\documentclass[aoas]{imsart}

\RequirePackage{amsthm,amsmath,amsfonts,amssymb}
\RequirePackage[authoryear]{natbib}
\RequirePackage{algorithm}
\RequirePackage{algorithmicx}
\RequirePackage{algpseudocode}
\RequirePackage{caption}
\RequirePackage{subcaption}
\RequirePackage{graphicx}
\usepackage{xcolor}
\usepackage{natbib}
\startlocaldefs
\theoremstyle{remark}
\newtheorem{remark}{Remark}
\newcommand{\E}{E}
\newcommand{\I}{I}
\renewcommand{\Pr}{\mathrm{pr}}

\newcommand{\Cov}{\mathrm{cov}}

\endlocaldefs

\begin{document}

\begin{frontmatter}
\title{Bootstrapping the Cross-Validation Estimate}
\runtitle{Bootstrapping the Cross-Validation Estimate}

\begin{aug}
\author[A]{\fnms{Bryan}~\snm{Cai}\ead[label=e1]{bxcai@stanford.edu}}
\author[B]{\fnms{Yuanhui}~\snm{Luo}}
\author[B]{\fnms{Xinzhou}~\snm{Guo}}
\author[C]{\fnms{Fabio}~\snm{Pellegrini}}
\author[C]{\fnms{Menglan}~\snm{Pang}}
\author[C]{\fnms{Carl}~\snm{de Moor}}
\author[C]{\fnms{Changyu}~\snm{Shen}}
\author[D]{\fnms{Vivek}~\snm{Charu}}
\and
\author[E]{\fnms{Lu}~\snm{Tian}}
\address[A]{Department of Computer Science, Stanford University\printead[presep={,\ }]{e1}}
\address[B]{Department of Mathematics, The Hong Kong University of Science and Technology}
\address[C]{Biogen Inc}
\address[D]{Department of Medicine, Stanford University}
\address[E]{Department of Biomedical Data Science, Stanford University}
\end{aug}

\begin{abstract}
Cross-validation is a widely used technique for evaluating the performance of prediction models, ranging from simple binary classification to complex precision medicine strategies. It helps correct for optimism bias in error estimates, which can be significant for models built using complex statistical learning algorithms. However, since the cross-validation estimate is a random value dependent on observed data, it is essential to accurately quantify the uncertainty associated with the estimate. This is especially important when comparing the performance of two models using cross-validation, as one must determine whether differences in estimated error are due to chance. Although various methods have been developed to make inferences on cross-validation estimates, they often have many limitations, such as requiring stringent model assumptions. This paper proposes a fast bootstrap method that quickly estimates the standard error of the cross-validation estimate and produces valid confidence intervals for a population parameter measuring average model performance. Our method overcomes the computational challenges inherent in bootstrapping a cross-validation estimate by estimating the variance component within a random-effects model. It is also as flexible as the cross-validation procedure itself. To showcase the effectiveness of our approach, we conducted comprehensive simulations and real-data analysis across two applications. 
\end{abstract}


\end{frontmatter}

\section{Introduction}

Predictive modeling has emerged as a prominent tool in biomedical research, encompassing diverse applications such as disease diagnosis, patient risk stratification, and personalized treatment recommendations \citep{sullivan2004presentation, hemann2007framingham,  solomon2006renal,krittanawong2017artificial}. A wide range of methods have been employed to create prediction models, from basic linear regression to sophisticated deep learning algorithms. Once these models are developed, it is crucial to assess their performance for several reasons. First, the output of a model cannot be effectively utilized or interpreted without understanding its accuracy. 
Second, with a wealth of prediction tools at hand, choosing the best model from a set of candidates can be challenging, with multiple factors affecting the decision, including, but not limited to, cost, interpretability, and the model's performance in an external population. Lastly, even in the model construction phase, evaluating the model performance is often needed for fine-tuning. For example, when applying neural networks, the network structure needs to be specified by the analyst to optimize prediction performance.

The performance of prediction models can be measured in various ways, depending on the intended application. If the model aims to predict a continuous or binary outcome of interest, its accuracy can be measured by the mean absolute prediction error or receiver operating characteristic (ROC) curve, respectively. Some settings are more complex, such as evaluating how good an empirically derived precision medicine strategy is. Specifically, novel therapies are often found to benefit some patients more than others. This treatment effect heterogeneity suggests possible implementation of precision medicine strategies that recommend appropriate treatments for different patients, maximizing the benefit for each individual. The performance of such a recommendation rule can be measured by the conditional average treatment effect (CATE) among patients who are recommended to receive a particular treatment according to the recommendation. Our motivating example is the clinical trial “Prevention of Events with Angiotensin Converting Enzyme Inhibition” (PEACE) \citep{peace2004angiotensin}.  PEACE trial randomized 8,290 patients to receive either an ACE inhibitor (ACEi) or placebo to examine the potential effect of ACEi on reducing cardiovascular risk for patients with stable coronary artery disease and normal or slightly reduced left ventricular function. It was inconclusive whether ACEi therapy could reduce mortality in the entire study population with the estimated hazard ratio (ACEi vs placebo) of 0.92 (95\% confidence interval: 0.78 to 1.08, $p=0.30$).  However, ACEi has been reported to still be effective in reducing future cardiovascular risk in selected patients \citep{solomon2006renal}. It is desirable to identify a high-value subgroup of patients who may benefit from ACEi. To this end, one can build a scoring system to capture the individualized treatment effect, based on which a recommendation on the use of ACEi can be made for individual patients. In this case, it is important to assess the actual benefit in the patient subgroup consisting of those who were recommended to receive ACEi.

Cross-validation is a commonly used technique to assess the performance of a predictive model and overcome the over-optimistic bias that results from using the same data for both training and evaluation \citep{efron1997improvements}. The approach involves splitting observed data into a training set and a testing set, with only the latter being used to evaluate the performance of the model trained based on the former, thus avoiding the optimism bias. In our motivating example, it involves deriving the precision medicine strategy from the training set and estimating the CATE in the subgroup of patients, who are recommended to receive ACEi in the testing set. In general, the resulting cross-validation estimator is a random quantity, dependent on both the random splitting of the data and the observed data itself. To reduce the randomness due to the former cause, one can repeat the training and testing process multiple times and average the prediction performance on the testing data. The randomness inherent in the observed data, however, reflects the fact that if a new set of data was sampled from the underlying population, the cross-validation results would be different from the current one. In essence, the cross-validation estimate is a statistic, or a function of observed data, despite its complex construction, and thus a random variable itself. 
In the PEACE example, we derived a precision medicine strategy recommending ACEi use from a training set consisting of 80\% of the study population and evaluated its performance on a testing set consisting of 20\% of the study population. Based on 500 cross-validations, the cross-validated estimate for CATE in terms of the difference in restricted mean survival time (RMST) among patients who are recommended to receive ACEi was 21.1 days favoring ACEi. In contrast, the cross-validated estimate for the difference in RMST among patients who are recommended not to receive ACEi was -13.2 days favoring placebo. While these results suggest a potential value of the derived treatment recommendation rule, their statistical significance after accounting for the estimation uncertainty remains unknown.  

Realizing this fact, it is important to derive and estimate the distribution of this statistic so that we can (1) understand the population parameter the cross-validation procedure estimates and (2) attach an appropriate uncertainty measure to the cross-validation estimate \citep{bayle2020cross, lei2020cross, yousef2021estimating}.  For the simple case of large sample size and small number of parameters,  the asymptotic distribution of the cross-validation estimator has been studied in depth \citep{dudoit2005asymptotics, tian2007model, 
ledell2015computationally}.  For example, when model training and validation are based on the same loss function, the cross-validation estimator is asymptotically Gaussian \citep{dudoit2005asymptotics}.  A computationally efficient variance estimator for the cross-validated area under the ROC curve has also been proposed, when the parameters used in the classifier are estimated at the root $n$ rate  \citep{ledell2015computationally}. More recently, Bayle et al. have established the asymptotic normality of general $K$-fold cross-validated prediction error estimates and proposed a consistent variance estimator under a set of stability conditions \citep{bayle2020cross}. The learning algorithm can be general and flexible, but the error estimate in the validation set needs to be in the form of a sum of identically independent distributed (i.i.d.) random elements. The validity of their proposed variance estimate requires that the variation from model training is dominated by that in estimating the prediction error based on testing data.  Following a similar line, Austern and Zhou (2020) have also studied the asymptotic distribution of the $K$ fold cross-validation estimator, allowing $K$ to increase with the sample size \citep{austern2020asymptotics}.  The regularity conditions for studying the $K$ fold cross-validation estimate proposed by \citet{austern2020asymptotics} are substantially more general than those in previous work, but can be difficult to verify in specific applications.  Furthermore, the proposed variance estimator relies on a nested cross-validation procedure with leave-one-out cross-validation as the bottom layer and can be very difficult to compute for some applications. A more computationally manageable nested cross-validation method has been proposed recently to automatically quantify the uncertainty of the cross-validation estimate and construct confidence intervals for the model performance \citep{bates2021cross}. The key is to use an additional loop of cross-validations to correct the finite-sample bias of the variance estimator proposed in \citet{bayle2020cross} and \citet{austern2020asymptotics}. However, this method still requires specific forms for the performance measure of interest.

In summary, the majority of previous work on cross-validation assumes a simple form for the performance measure, such as the average of a set of random variables, and sometimes requires that the prediction model be trained using the same loss function. Furthermore, the validity of the proposed confidence intervals (CIs) often relies on suitable stability conditions and large sample approximations. However, there are many applications of cross-validation that are not covered by these conventional approaches. For instance, currently there is no method that is directly applicable to cross-validation in evaluating the performance of a precision medicine strategy, e.g., in the aforementioned PEACE trial, with theoretical guarantees.  If we want to make inferences on the average treatment effect (ATE) among patients recommended for ACEi treatment, a new approach is required. Resampling methods are a general approach for estimating the variance of a statistic and can provide fairly accurate CIs with minimum model assumptions, even in small to moderate sample sizes. The main challenge in this context is computational cost, particularly when a complex and time-consuming algorithm is used to train the prediction model.  This paper seeks to address these limitations by characterizing the underlying population parameter estimated through the cross-validation procedure and proposing a general, computationally efficient resampling method for constructing CIs of this parameter. We aim to reduce the restrictions imposed by traditional approaches while maintaining accuracy and computational feasibility.

\section{Method}
In a very general setup, we use the random vector $X$ to denote individual observation, and the observed data consist of $n$ i.i.d. copies of $X$, i.e., $D_n=\{X_1,\cdots, X_n\}.$ The output of the training procedure is a parameter estimate, which can be a finite-dimensional vector or infinite-dimensional function, denoted by $\widehat{\psi}(D_n)$ to emphasize its dependence on observed data $D_n$ and the fact that it is a random quantity. In evaluating the ``performance'' of $\widehat{\psi}(D_n)$ in a new testing set consisting of $N$ i.i.d. observations $\tilde{D}_N=\{\tilde{X}_1, \cdots, \tilde{X}_N\},$ a summary statistic is calculated as a function of the testing data and $\widehat{\psi}(D_n),$ which can be written as $L\left\{\tilde{D}_N, \widehat{\psi}(D_n)\right\}.$ It is possible to make inferences and derive a confidence interval on this quantity by treating $\tilde{D}_N$ or both $D_n$ and $\tilde{D}_N$ as random. However, in most applications, we only have a single dataset, and the cross-validation procedure is needed to objectively evaluate the model performance. Specifically, in cross-validation, we randomly divide the observed data $D_n$ into two non-overlapping parts denoted by $D_{train}$ and $D_{test}$, and calculate 
$L\left\{D_{test}, \widehat{\psi}(D_{train})\right\}.$ In order to reduce the variability of random splits, the aforementioned step is oftentimes repeated many times and the average performance measure is obtained as the final cross-validation estimator of the model performance:
$$ \widehat{Err}^{CV}=\frac{1}{B_{CV}}\sum_{b=1}^{B_{CV}} L\left\{D_{test}^b, \widehat{\psi}(D_{train}^b)\right\},$$
where $D_n=D_{train}^b\cup D_{test}^b$ represents the $b$th split. The number of replications, $B_{CV}$, should be relatively large to reduce the Monte Carlo variation due to random splits. It is often in the range of several hundreds in practice. Note that although we used $Err$ to represent the model performance in consistency with notations used in the literature \citep{bates2021cross}, the performance measure is not necessarily a prediction error. Other metrics such as those discussed in the PEACE trial can also be used.

Many cross-validation applications can fit into this very general framework.  In this paper, we will focus on several typical examples. Due to the space limit, we presented two applications in the main paper. The third application for the cross-validated mean absolute prediction error can be found in Section 2 of the Supplementary Material \citep{cai2025bootstrapping}. 

\subsection{Application 1}
In our motivating example, we are interested in developing a precision medicine strategy and evaluating its performance in a randomized clinical trial setting. Specifically, the precision medicine strategy here is a binary classification rule to recommend a treatment to a patient based on his or her baseline characteristics to maximize the treatment benefit. 
Before applying this recommendation to clinical practice, it is important to estimate the uncertainty of the treatment effect in the identified subgroup recommended for treatment, to ensure that the anticipated stronger treatment effect is real. There are many different ways to construct such a treatment recommendation classifier \citep{chen2017general, tian2014simple}.  For example, one may first construct an individualized treatment response (ITR) score  by minimizing a loss function based on a training dataset $D_{train},$ 
\begin{equation}\frac{1}{m}\sum_{X_i\in D_{train}} \left\{Y_i-\gamma'\tilde{Z}_i-(G_i-\pi)\beta'\tilde{Z}_i \right\}^2 ,
\label{eq:lossITR}
\end{equation}
where $X_i=(Z_i,G_i,Y_i),$ $Y_i$ is the response of interest with a higher value being desirable, $Z_i$ is the baseline covariate, $\tilde{Z}_i$ is its counterpart including an intercept, $G_i \in \{0, 1\}$ is a binary treatment indicator and independent of $Z_i$ (i.e., the treatment is randomly assigned to patients in the training set), and $\pi=\Pr(G_i=1).$   Let $\widehat{\gamma}(D_{train})$ and $\widehat{\beta}(D_{train})$ be the minimizer of (\ref{eq:lossITR}). Then  $\widehat{\gamma}(D_{train})'\tilde{z}$ and $\widehat{\beta}(D_{train})'\tilde{z}$ can be used to approximate the conditional average outcome $E(Y|Z=z)$ and CATE, 
$$\Delta(z)=E\left(Y^{(1)}-Y^{(0)} \mid Z=z\right),$$ 
respectively \citep{tian2014simple, yadlowsky2021estimation}, where $\tilde{z}=(1, z')',$ $Y^{(g)}$ is the potential outcome if the patient receives treatment $g \in \{0, 1\},$ and the observed outcome $Y=Y^{(1)}G+Y^{(0)}(1-G).$  The intuition is that we may decompose the limit of the loss function (\ref{eq:lossITR}) as
\begin{align*}
\E\left[ \left(Y-\gamma'\tilde{Z} \right)^2\right] +\pi(1-\pi) E\left[\left(Y^{(1)}-Y^{(0)}-\beta'\tilde{Z}\right)^2\right]+const,
\end{align*}
and minimizing the original loss function with respect to $\beta$ amounts to minimizing 
$$ \E\left[\left\{\Delta(Z)-\beta'\tilde{Z} \right\}^2\right]$$
and approximating CATE $\Delta(z)$ with $\beta'\tilde{z}.$  Thus, the constructed ITR score is $\widehat{\Delta}(z\mid D_{train})=\widehat{\beta}(D_{train})'\tilde{z}$, which can be used to guide the treatment recommendation for the individual patient.  Once an estimated ITR score is available, treatment $G=1$ can be recommended to patients whose $\widehat{\Delta}(Z\mid D_{train})>c_0$ and treatment $G=0$ to patients whose $\widehat{\Delta}(Z\mid D_{train})\le c_0,$ where $c_0$ is a constant reflecting the ``cost'' of the treatment. Here, we choose $c_0=0$ for simplicity.  In the testing set, one may evaluate the performance of this recommendation rule by estimating the ATE in the subgroup of patients recommended to receive the treatment $G=1,$ i.e, $\widehat{D}_{test}^{(1)}=\left\{X \in D_{test} \mid \widehat{\Delta}(Z\mid D_{train})>0\right\}$ and among the subgroup of patients recommended to receive the treatment $G=0,$ i.e., $\widehat{D}_{test}^{(0)}=\left\{X \in \widehat{D}_{test} \mid \widehat{\Delta}(Z\mid D_{train})\le 0\right\}.$  Specifically, we may consider the observed treatment effects
$$ \widehat{\Delta}_g(D_{train},D_{test})= \frac{\sum_{X_i\in \widehat{D}_{test}^{(g)}}Y_iG_i}{\sum_{X_i\in \widehat{D}_{test}^{(g)}} G_i}-\frac{\sum_{X_i\in \widehat{D}_{test}^{(g)}}Y_i(1-G_i)}{\sum_{X_i\in \widehat{D}_{test}^{(g)}} (1-G_i)}, g\in \{0, 1\}.$$

If $\widehat{\Delta}_1(D_{train}, D_{test})$ takes a ``large" positive value and $\widehat{\Delta}_0(D_{train}, D_{test})$ takes a ``large'' negative value, (in other words, the treatment effect is indeed estimated to be greater among those who are recommended to receive the treatment based on the constructed ITR score), then we may conclude that $\widehat{\Delta}(Z\mid D_{train})>0, $ is an effective treatment recommendation rule. 
In cross-validation, we may repeatedly divide the observed dataset $D$ into training and testing sets, $(D_{train}, D_{test}),$ and obtain the corresponding cross-validated treatment effect estimate $\widehat{\Delta}_g(D_{train}, D_{test}), g\in \{0, 1\}$.  In the end, the sample average of those resulting cross-validated treatment effect estimators is our final cross-validation estimator measuring the performance of the treatment recommendation system. In this application $X=(Z,G,Y)$ with $Z$, $G$ and $Y$ being predictors, the treatment assignment indicator, and a binary outcome, respectively, $\widehat{\psi}(D_{train})=\widehat{\beta}(D_{train}),$ and 
$$L\left(D_{test}, \psi\right)=\frac{\sum_{X_i\in D_{test}}Y_iG_i\I(\psi'Z_i>0)}{\sum_{X_i\in D_{test}} G_i \I(\psi'Z_i>0)}-\frac{\sum_{X_i\in D_{test}}Y_i(1-G_i)\I(\psi'Z_i>0)}{\sum_{X_i\in D_{test}}(1-G_i)\I(\psi'Z_i>0)}.$$

\subsection{Application 2}\label{sec:application2}
Beyond our motivating example, this general framework also covers other cross-validation applications. In the second example, we are interested in evaluating the performance of a prediction model in predicting binary outcomes by its c-index, which is the area under the ROC curve.
 The result can help us determine, for example, whether the c-index of a new prediction model is significantly higher than a desirable level. The prediction model can be constructed via fitting a logistic regression model, i.e., calculating a regression coefficient vector $\widehat{\beta}(D_{train})$ by maximizing the log-likelihood function,
$$ \sum_{(Z_i,Y_i)\in D_{train} } \left[ \beta'\tilde{Z}_i Y_i-\log\left\{1+\exp(\beta'\tilde{Z}_i)  \right\}\right],$$
based on a training dataset $D_{train},$ where $Z_i$ is the predictor, $\tilde{Z}_i=(1, Z_i')',$ and $Y_i\in \{0, 1\}$ is the binary outcome. If the dimension of $Z_i$ is high, lasso-regularization can be used in estimating $\beta$.  In any case, the c-index in a testing set $D_{test}$ can be calculated as 
$$\widehat{\theta}(D_{train}, D_{test})=\frac{1}{\tilde{n}_{test, 0}\tilde{n}_{test, 1}}\sum_{X_i \in D_{test}(0)}\sum_{X_j\in D_{test1}(1)}\I\left(\widehat{\beta}(D_{train})'\tilde{Z}_i<\widehat{\beta}(D_{train})'\tilde{Z}_j \right),$$
where $\tilde{n}_{test,g}$ is the number of observations in the set $D_{test}(g)=\{X_i=(Z_i,Y_i)\in D_{test}: Y_i=g\}, g\in \{0, 1\}.$
In cross-validation, we may repeatedly split the observed data $D_n$ into training and testing sets, $(D_{train}, D_{test}),$ and obtain the corresponding cross-validated c-indexes $\widehat{\theta}(D_{train}, D_{test})$.  In the end, the sample average of those resulting cross-validated c-index estimators is our final estimator measuring the predictive performance of the logistic regression. In this application, $X=(Z, Y)$ with $Z$ and $Y$ being the predictor and a binary outcome of interest, respectively, $\widehat{\psi}(D_{train})=\widehat{\beta}(D_{train}),$ and 
$$L\left(D_{test}, \psi\right)=\frac{1}{\tilde{n}_{test, 0}\tilde{n}_{test, 1}}\sum_{X_i\in D_{test}(0)}\sum_{X_j\in D_{test}(1)} \I\left(\psi'\tilde{Z}_i<\psi'\tilde{Z}_j \right).$$

\subsection{The Estimand of Cross-validation}
The first important question is what population parameter the cross-validation procedure estimates. As discussed in \citet{bates2021cross}, there are several candidates.  The first obvious population parameter is 
$$ Err(D_n)=  \lim_{N \rightarrow \infty} L\left(\tilde{D}_N, \widehat{\psi}(D_n)\right),$$
 where $D_n$ is the observed data of sample size $n$ and $\tilde{D}_N$ is a new independent testing set of sample size $N$ drawing from the same distribution as the observed data $D_n$. In practice, $\tilde{D}_N$ is oftentimes not available. This parameter depends on the observed data $D_n$, and directly measures the performance of the prediction model obtained from observed data $D_n$ in a future population. The second population parameter of interest is 
$$ Err_n=\E\{Err(D_n)\},$$
the average performance of prediction models trained based on ``all possible'' training sets of size $n$ sampled from the same distribution as the observed data $D_n.$ The subscript $n$ emphasizes the fact that this population parameter only depends on the sample size of the training set $D_n.$ While $Err(D_n)$ is the relevant parameter of interest in most applications, where we want to know the future performance of the prediction model at hand,  $Err_n$ is a population parameter reflecting the expected performance of prediction models trained via a given procedure. The prediction performance of the model from the observed data $D_n$ can be better or worse than this expected average performance. It is known that the cross-validation targets evaluating a training procedure rather than the particular prediction model obtained from the training procedure. Specifically, the cross-validation estimator $\widehat{Err}^{CV}_m$ actually estimates $Err_m$ in the sense that 
$$\left(\begin{array}{c} \widehat{Err}_m^{CV} \\ Err(D_n) \end{array}\right)\approx \left(\begin{array}{c} Err_m+\epsilon \\ Err_n+\zeta \end{array}\right),$$
where $m$ is the sample size of the training set used in the cross-validation, i.e., the dataset $D_n$ is divided into a training set of size $m$ and a testing set of size $(n-m)$ in cross-validation. Here, $\epsilon$ and $\zeta$ are two mean zero random noises and oftentimes approximately independent.
In many cases $Err_n\approx Err_m$, when $m$ is not substantially smaller than $n.$ If we ignore their differences, then $\widehat{Err}^{CV}_m$ can also be viewed as an approximately unbiased estimator for $Err(D_n)$, because 
$$\widehat{Err}^{CV}_m-Err(D_n)=(Err_m-Err_n)+(\epsilon-\zeta),$$ 
whose expectation is approximately zero. On the other hand, the variance of $\widehat{Err}^{CV}_m-Err(D_n)$ tends to be substantially larger than the variance of $\widehat{Err}^{CV}_m-Err_m,$ since $\epsilon$ and $\zeta$ are often independent and the variance of $\zeta$ is nontrivial relative to that of $\epsilon$. This is analogous to the phenomenon that the sample mean of observed data is a natural estimator of the population mean. It also can be viewed as an unbiased ``estimator'' of the sample mean of a set of future observations, because the expectation of sample mean of future observations is the same as the population mean, which can be estimated by the sample mean of observed data.  In this paper, we take $Err_m$ as the population parameter of interest, because approximately $Err(D_n)$ is simply $Err_m$ plus a random noise $\zeta$, which may be independent of the cross-validation estimate.  In other words, we take the view that the cross-validation estimate evaluates the average performance of a training procedure rather than the performance of a particular prediction model.  As the sample size goes to infinity, we write $Err=\lim_{n\rightarrow \infty}Err_n.$ When $n$ is sufficiently large, $Err\approx Err_n \approx Err(D_n).$  

\subsection{A Toy Example for The Estimand of Cross-validation}\label{sec:toyexample}
\begin{figure}
\centering
\begin{subfigure}[b]{0.45\linewidth}
\centering
\includegraphics[width=\linewidth]{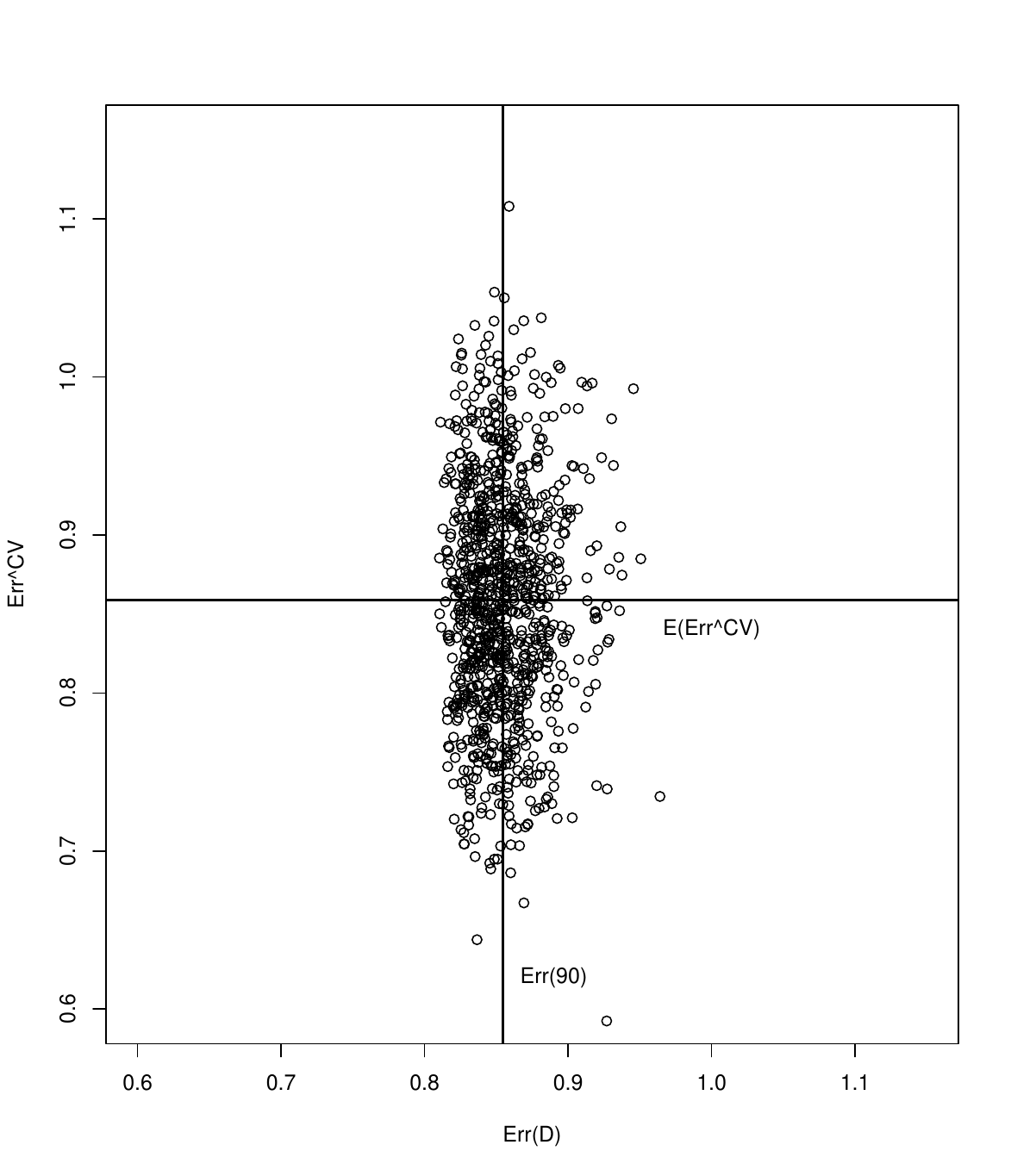}
\caption{$Err(D_n)$ vs $\widehat{Err}^{CV}_m$ for estimating the mean absolute prediction error}
\label{fig:errscatter}
\end{subfigure}
\hfill
\begin{subfigure}[b]{0.45\linewidth}
\centering
\includegraphics[width=\linewidth ]{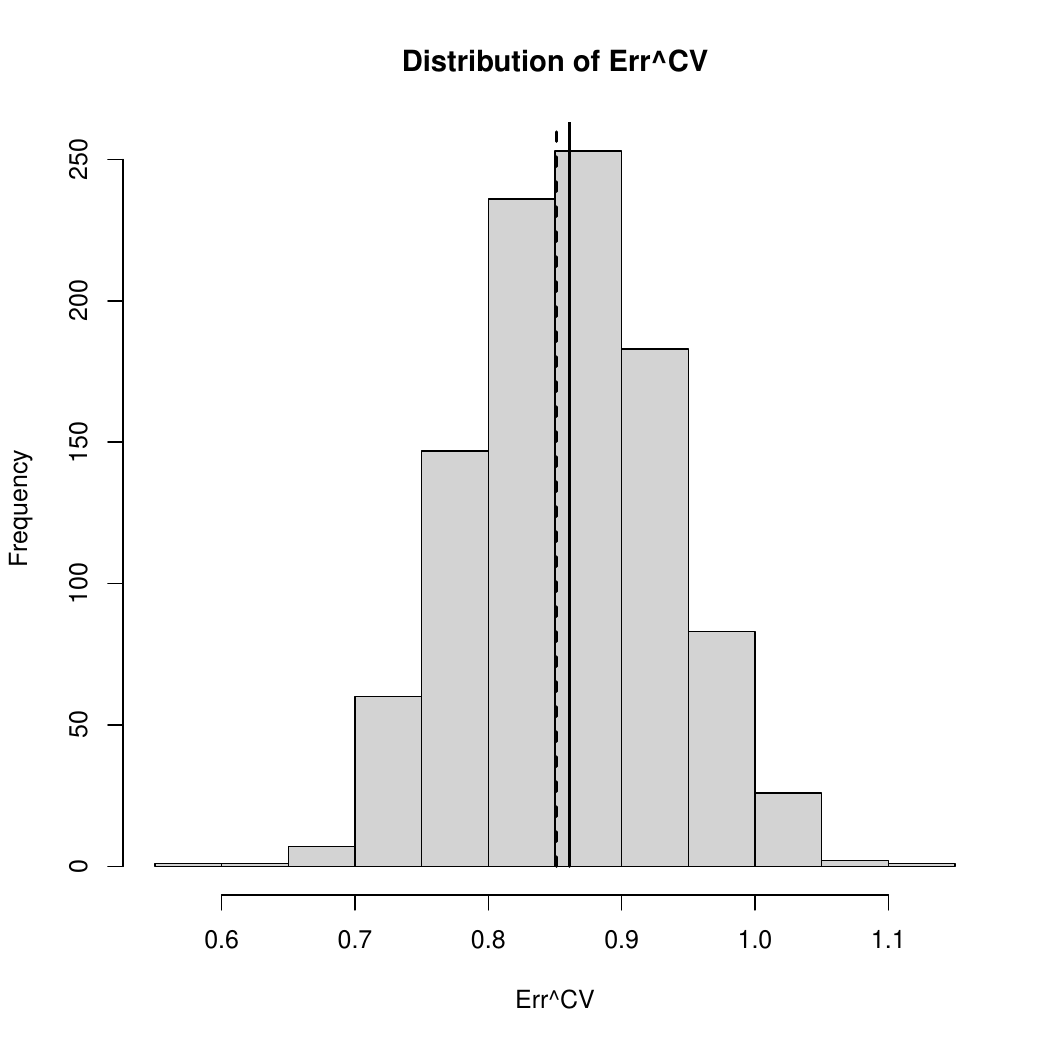}
\caption{The distribution of $\widehat{Err}^{CV}_m$ for estimating the mean absolute prediction error; $Err_n$: dashed line; $Err_m$: solid line}
\label{fig:errdist}
\end{subfigure}
\end{figure}

In the following section, we use a simple example to demonstrate the relationship between the various parameters of interest. 
Suppose that covariate $Z_i\sim N(0, I_{10})$,  and the response
$Y_i=\alpha_0+\beta_0'Z_i+\epsilon_i, i=1,\cdots, n,$
where $I_{10}$ is a 10 by 10 identity matrix, $\alpha_0=0$, $\beta_0=(1, 1, 1, 1, 0, 0, 0, 0, 0, 0)',$  $\epsilon_i\sim N(0, 1)$ and $n=90.$  We were interested in constructing a prediction model via fitting a linear regression model and evaluating its performance in terms of the mean absolute prediction error. To this end, for each simulated dataset $D_n=\{X_i=(Z_i, Y_i), i=1,\cdots, n\},$  we estimated the regression coefficients of the linear model by ordinary least squares method and denoted the estimators of $\alpha_0$ and $\beta_0$ by $\widehat{\alpha}(D_n)$ and $\widehat{\beta}(D_n),$ respectively.  Then we calculated the true mean absolute prediction error as the expectation of $|G|,$ where $G\sim N\left(\widehat{\alpha}(D_n), 1+\|\widehat{\beta}(D_n)-\beta_0\|_2^2\right)$ is a random variable.   This expectation was $Err(D_n)$, the prediction error of the model trained based on dataset $D_n$ in a future population.  We also constructed the cross-validation estimate of the prediction error by repeatedly splitting $D_n$ into a training set of size $m=80$ and a testing set of size $n-m=10.$  The resulting estimator for the estimation error was $\widehat{Err}_m^{CV}.$  Repeating these steps, we obtained 1000 pairs of $Err(D_n)$ and $\widehat{Err}_m^{CV}$ from simulated datasets. The empirical average of 1000 $Err(D_n)$s was an approximation to $Err_n=\E\left\{Err(D_n)\right\}.$  Figure \ref{fig:errscatter} is a scatter plot of $Err(D_n)$ vs. $\widehat{Err}^{CV}_m$ based on 1000 simulated datasets. It was clear that $\widehat{Err}^{CV}_m$ and $Err(D_n)$ were almost independent but shared a similar center. Specifically, the empirical average of $\widehat{Err}^{CV}_m$ was 0.859, and the empirical average of $Err(D_n)$ was 0.851. Therefore, the cross-validated error estimator $\widehat{Err}^{CV}_m$ can be viewed as a sensible estimator for $Err_n=\E\left\{Err(D_n)\right\}\approx 0.851$, and more precisely an unbiased estimator for $Err_m$, whose value was estimated as 0.861 using the same simulation described above. Note that $(m, n)=(80, 90)$ and $n$ and $m$ are fairly close. The distribution of the cross-validated error estimators along with $Err_m$ and $Err_n$ is plotted in Figure \ref{fig:errdist}, suggesting that the small difference between $Err_{80}$ and $Err_{90}$ is negligible relative to the variation of the cross-validation estimator $\widehat{Err}^{CV}_m$ itself. In addition, $\widehat{Err}^{CV}_m$ can also be thought as a ``prediction'' to $Err(D_n),$ since the latter was approximately $Err_n$ plus an independent mean zero normally distributed ``measurement error''.

\subsection{Statistical Inferences on $Err_m$}
In this section, we aim to construct a valid CI for $Err_m$ based on the cross-validated estimate.  First, we define the cross-validated estimate with repeated training and testing splits as 
$$ \widehat{Err}^{CV}_m = \E\left[L\left\{D_{test}^b, \widehat{\psi}(D_{train}^b)\right\}\right]$$
where the expectation is with respect to random division of training and testing sets. One may anticipate that $\widehat{Err}^{CV}_m$ is a ``smooth" functional with respect to the empirical distribution of observed data $D_n$, because each individual observation's contribution to the final estimator is ``averaged'' across different training and testing divisions.  Therefore, we expect that $\widehat{Err}^{CV}_m$ is a root-$n$ regular estimator of $Err$ and $Err_m$, i.e.,
$$\sqrt{n}\left(\widehat{Err}^{CV}_m-Err_m\right) \rightarrow N(0, \sigma_0^2),$$
in distribution as $n$, the sample size of $D_n$, goes to infinity and $\lim_{n\rightarrow \infty} m/n \in (0, 1).$  In Section 1 of the Supplementary Material \citep{cai2025bootstrapping}, we have provided a set of sufficient conditions for the aforementioned large sample approximation.  
In such a case, an asymptotic CI for $Err_m$ can be constructed as 
$$ \left[\widehat{Err}^{CV}_m-1.96 \frac{\widehat{\sigma}}{\sqrt{n}},  \widehat{Err}^{CV}_m+1.96 \frac{\widehat{\sigma}}{\sqrt{n}} \right],$$
where $\widehat{\sigma}$ is a consistent estimator of the standard error $\sigma_0.$
However, in general, it is difficult to obtain such a consistent variance estimator when complex procedures such as lasso regularized regression or random forest are used to construct the prediction model as in the two application examples discussed above. An appealing solution is to use the non-parametric bootstrap  described in algorithm \ref{ag:NB} to estimate $\sigma_D^2$ \citep{davison1997bootstrap, efron1994introduction}. The rationale is that, under the same set of assumptions in Section 1 of the Supplementary Material \citep{cai2025bootstrapping}, 
$$\sqrt{n} \left(\begin{array}{c} 
\widehat{Err}^{CV}_m-Err_m\\
\widehat{Err}^{CV*}_m-Err_m \end{array} \right)=\frac{1}{\sqrt{n}}\sum_{i=1}^n \left(\begin{array}{c}\tau(X_i)\\ \tau(X_i)W_i  \end{array} \right)+o_{P^*}(1),$$
where $\widehat{Err}^{CV*}_m$ is the cross-validated estimator based on bootstrapped data $D^*_n$, $(W_1, \cdots, W_n)\sim Multn\left(n, (1/n, \cdots, 1/n)\right)$, $W_i$ is the number of observation $X_i$ in $D^*_n,$ $\tau(X_i)$ is a mean zero influence function associated with the cross-validation estimate, and $P^*$ is the product probability measure with respect to random data and the independent random weights $(W_1,\cdots, W_n).$ Therefore, conditional on observed data $D_n,$
$$\sqrt{n}\left(\widehat{Err}^{CV*}_m-\widehat{Err}^{CV}_m\right)=\frac{1}{\sqrt{n}}\sum_{i=1}^n \tau(X_i)(W_i-1)+o_{P^*}(1),$$
converges weakly to a mean zero Gaussian distribution with a variance of 
$$\frac{1}{n}\sum_{i=1}^n\tau(X_i)^2 \rightarrow \sigma_0^2,~~\mbox{as}~~ n\rightarrow \infty.$$

\begin{remark}
 The conditions in Section 1 of the Supplementary Material \citep{cai2025bootstrapping} can be significantly relaxed, while the distribution of the cross-validation estimate can still be approximated by the corresponding bootstrapped distribution. The intuition is that bootstrap methods can be used to estimate the variance of a ``regular'' statistic, and the cross-validation estimate is often ``regular'' because the evaluation function $L(D, \psi)$ is relatively simple and ``well behaved''.  For example, in our numerical study, we have found that our proposed bootstrap method performs well in evaluating the performance of a random forest or lasso-regularized regression model, although those models are not ``regular'' and do not strictly satisfy the regularity conditions given in the Supplementary Material \citep{cai2025bootstrapping}. The details of the corresponding simulation study can be found in Section 5 of the Supplementary Material \citep{cai2025bootstrapping}.  
However, the focus of this paper is not to explore the most general sufficient conditions for the validity of the bootstrap method but to propose a computationally efficient resampling method to estimate $\sigma_D^2$ and construct a CI for $Err_m$ in practice.
\end{remark}

Operationally, Algorithm \ref{ag:NB} is expected to generate a consistent variance estimator of $\sigma_D^2$, $\widehat{\sigma}_D^2.$
\begin{algorithm}[ht]
  \caption{Naive Bootstrap}\label{ag:NB}
  {\small
  \begin{algorithmic}[1]
      \For{$b \gets 1$ to $B_{BOOT}$} 
        \State Sample original data to form a bootstrapped dataset of the size $n$ denoted by $D_b^*;$
        \State Perform cross-validation based on bootstrapped dataset $D_b^*;$
         \For{$k \gets 1$ to $B_{CV}$} 
        \State Randomly split $D^*_b$ into $D^{*(k)}_{b,train}$ of size $m$ and $D^{*(k)}_{b, test}$ of size $n-m;$
        \State Calculate $L\left\{D_{b, test}^{*(k)}, \widehat{\psi}(D_{b, train}^{*(k)})\right\}$
      \EndFor
      \State      Calculate the bootstrapped cross-validation estimate 
$$\widehat{Err}_{b,m}^{CV*}=\frac{1}{B_{CV}}\sum_{k=1}^{B_{CV}}L\left\{D_{b, test}^{*(k)}, \widehat{\psi}(D_{b, train}^{*(k)})\right\}.$$
      \EndFor
\end{algorithmic}
}
\end{algorithm}
The variance $\sigma_D^2$ can be estimated by $n$ times the empirical variance of $B$ bootstrapped cross-validation estimates $\left\{\widehat{Err}_{1,m}^{CV*}, \cdots, \widehat{Err}_{B_{BOOT},m}^{CV*}\right\}.$  However, there are several concerns in this naive resampling procedure, which may result in poor performance in practice. 
\begin{itemize}
\item The bootstrap procedure samples observations with replacement and results in potential duplicates of the same observation in the bootstrapped dataset. Naively splitting the bootstrapped dataset into training and testing sets results in an overlap between them, which can induce optimistic bias in evaluating the model performance. If we apply the naive bootstrap method to analyze the Toy Example described in Section 2.4, then the empirical average of bootstrapped cross-validation estimates $\widehat{Err}_{b, m}^{CV*}$ was downward biased in comparison with  $Err_m^{CV}$ by 0.80 standard deviation of cross-validation estimates $\widehat{Err}_m^{CV}.$ 
\item The training set of size $m$ in the bootstrapped dataset $D^*_n$ contains substantially fewer than $m$ distinct observations, which reduces the ``effective" sample size for training a prediction model and induces a downward bias in evaluating the average model performance. This downward bias may be smaller or greater than the optimism bias induced by the overlap between training and testing sets depending on specific applications, but is undesirable. 
\item More importantly, to obtain a cross-validated estimate for each bootstrap sample,  one needs to perform the cross-validation multiple times to reduce the Monte Carlo variation due to random training/testing divisions, e.g., $B_{CV} \ge 200.$  In addition, the number of bootstraps also cannot be too small. The conventional recommendation for estimating the variance of a statistic using bootstrap is to let $B_{BOOT}= 400 - 1000.$ In such a case, one needs to train and evaluate the prediction model $>80,000$ times and the corresponding computational burden can be prohibitive for complex training algorithms.  
\end{itemize}

In this paper, we present a modified bootstrap procedure to overcome aforementioned difficulties.  First, in implementing cross-validation on a bootstrapped dataset, we view bootstrapped data as weighted samples of the original data, i.e., observation $X_i$ is weighted by a random weight $W_i$, which is the number of this observation selected in this bootstrap iteration. In cross-validation, we first split the original dataset into training and testing sets, $D_n=D_{train}\cup D_{test},$ and bootstrapped training and testing sets denoted by $D^*_{train}$ and $D^*_{test}$, respectively, are then constructed by collecting all observations in $D_{train}$ and $D_{test},$ respectively, but with their respective bootstrap weights. Since $D_{train}$ and $D_{test}$ have no overlap, $D^*_{train}\cap D^*_{test}=\phi$ as well. Therefore, we don't allow the same observation to be used in both training and testing. One consequence is that the sample sizes of $D^*_{train}$ and $D^*_{test}$ are not fixed across different bootstrap samples.  But their average sample sizes remain the same as those of $D_{train}$ and $D_{test}.$  

Second, we note that the effective sample size of the training set based on the bootstrapped data can be substantially smaller than $m$. Specifically, the number of distinct observations in $D_{train}^*$ is $0.632m$ on average \citep{efron1997improvements}.  Therefore, it is desirable to increase the number of distinct observations of $D_{train}^*$ by allocating more observations to $D_{train}$, which is used to generate $D_{train}^*$ in a bootstrapped dataset. Ideally, we may want to increase the sample size of $D_{train}$ such that the number of distinct observations used for training is close to $m$ in bootstrapped cross-validation, which requires to increase the sample size in $D_{train}$ from $m$ to $m/0.632.$ On the other hand, the sample size of $D_{test}$ and thus $D_{test}^*$ would be reduced by using a larger training set in the bootstrapped cross-validation and such a large reduction in testing sample size may increase the variance of the resulting estimate. In summary, while we want to increase the sample size in the training set to reduce the bias of estimating the model performance in bootstrapped cross-validation due to the fact that fewer distinct observations are used to train the prediction model, we also want to limit the reduction in the number of testing samples so that the variance of the cross-validation estimate would not be greatly affected by this adjustment in training and testing sample sizes. A compromise is to find an adjusted sample size $m_{adj}$ by minimizing the loss function 
\begin{equation}
\left(\frac{m_{adj}}{m/0.632}-1\right)^2+\lambda_0 \left(\frac{n-m}{n-m_{adj}}-1\right)^2,  \label{eq:sizeadj}
\end{equation}
where the first and second terms control the closeness of the ``effective" sample size in the bootstrapped training set to $m$ and the relative change in the sample size of the testing set after the adjustment, respectively. Here, $\lambda_0$ controls the relative importance of these two tasks in determining the final adjustment. In our limited experience, we have found that the performance of the resulting resampling procedure was not very sensitive to the selection of this penalty parameter within a wide range, and we recommend setting $\lambda_0=1-0.632=0.368$ in practice.

More importantly, to alleviate the computational demand for the bootstrap procedure, we propose algorithm \ref{ag:mainboot}.
\begin{algorithm}[h!]
  \caption{Bootstrap Cross-Validation}\label{ag:mainboot}
  {\small
  \begin{algorithmic}[1]
      \For{$b \gets 1$ to $B_{BOOT}$} 
        \State Obtain a bootstrapped dataset $D^*_b$ and frequencies of all observations $\{W_{1b}, \cdots, W_{nb}\}.$
        \State  Calculate $m_{adj}$ by minimizing the loss function in (\ref{eq:sizeadj}) with $\lambda_0=0.368$;
         \For{$k \gets 1$ to $B_{CV}$} 
        \State Split $D_n$ into training and testing sets: $D^{(k)}_{b, train}$ of size $m_{adj}$ and $D^{(k)}_{b, test}$ of size $n-m_{adj};$
        \State Construct the training and testing sets $D_{b,train}^{*(k)}$ and $D_{b, test}^{*(k)}$ by weighing patients in $D^{(k)}_{b, train}$ and $D^{(k)}_{b, test}$ with their bootstrap weights $\left\{W_{ib}, i=1,\cdots, n\right\}.$
        \State Calculate the cross-validation error $ \theta^*_{bk}=L\left\{D^{*(k)}_{b, train}, D^{*(k)}_{b,test}\right\}$
      \EndFor      
      \EndFor

\State   Fit a random effects model \citep{
laird1982random} 
$$\theta^*_{bk}=\theta_0+\epsilon^*_{b}+\epsilon_{bk},  b=1, \cdots, B_{BOOT}; k=1, \cdots, B_{CV}$$
where $\epsilon^*_b\sim N(0, \sigma_{BT}^2)$ and $\epsilon_{bk}$ are independent mean zero noise with a variance of $\tau_0^2$.
\State Let
$$\widehat{\sigma}_{BT}^2=\frac{1}{B_{BOOT}-1}\sum_{b=1}^{B_{BOOT}} (\bar{\theta}^*_{b}-\bar{\theta}^*)^2-\frac{\widehat{\tau}_0^2}{B_{CV}},$$
where $\bar{\theta}_b^*=B_{CV}^{-1}\sum_{k=1}^{B_{CV}} \theta^*_{bk},$ $\bar{\theta}^*=B_{BOOT}^{-1}\sum_{b=1}^{B_{BOOT}} \bar{\theta}^*_b,$  and 
$$\widehat{\tau}_0^2=\frac{1}{(B_{CV}-1)B_{BOOT}}\sum_{b=1}^{B_{BOOT}} \sum_{k=1}^{B_{CV}}(\theta^*_{bk}-\bar{\theta}^*_{b})^2$$ is an estimator for $\tau_0^2.$ \label{line1:bootest}
\State $$\widehat{\sigma}_m^{CV} \gets \widehat{\sigma}_{BT}$$ 
is our bootstrap standard error estimator of cross-validation estimator $\widehat{Err}^{CV}_m$ and the 95\% CI for $Err_m$ is 
$$\left[\widehat{Err}^{CV}_m-1.96\times \widehat{\sigma}_m^{CV},  \widehat{Err}^{CV}_m+1.96\times \widehat{\sigma}_m^{CV} \right].$$ \label{line2:bootest}
\end{algorithmic}
}
\end{algorithm}
The rationale is that the center of $\theta^*_{bk}=L\left\{D^{*(k)}_{b, train}, D^{*(k)}_{b,test}\right\}$, $(\theta_0+\epsilon_{b}^*),$ is approximately the cross-validation estimate based on the bootstrapped dataset $D^*_b$ as the number of random training and testing divisions increasing to infinity. Under this framework, $\sigma_{BT}^2$ measures the between-bootstrap variance, which is the bootstrap variance estimator we aim to calculate,  and $\tau_0^2=\mbox{var}(\epsilon_{bk})$ measures the within-bootstrap variance, i.e., the variance due to random training and testing divisions. The empirical variance of $\left\{\bar{\theta}^*_1, \cdots, \bar{\theta}^*_B\right\}$ based on a very large $B_{CV}$ is approximately unbiased in approximating $\sigma_{BT}^2,$ corresponding to the naive bootstrap procedure.  However, this naive approach is very inefficient and there is no need to choose a very large $B_{CV}$ to eliminate all Monte Carlo variance in estimating the cross-validation prediction error for every bootstrapped dataset as in algorithm \ref{ag:NB}. Alternatively, a good moment estimate for the variance component in the random effects model can be constructed with a moderate $B_{CV}$, say $10-20,$ and a reasonably large $B_{BOOT},$ say 400. This can substantially reduce the computational burden from 80,000 model training to 8,000 model training. We have conducted comprehensive numerical studies to examine the real time saving of algorithm 2 in comparison with naive bootstrap under various practical settings and successfully confirmed the theoretical gain in computational efficiency. The detailed results are reported in Section 5 of the Supplementary Material \citep{cai2025bootstrapping}.

\begin{remark}
The total number of model training is $B_{BOOT}\times B_{CV}.$  A natural question is how to efficiently allocate the number of bootstraps and number of cross-validations per bootstrap given the total number of model training.   The variance estimator, $\widehat{\sigma}_{BT}^{2},$ is a random statistic itself with a variance \citep{boardman1974confidence, williams1962confidence}, which can be approximated by 
$$  2\left(\frac{\left(\widehat{\sigma}_{BT}^2+B_{CV}^{-1}\widehat{\tau}_{0}^2\right)^2}{(B_{BOOT}-1)} \right)+2\left(\frac{\left(B_{CV}^{-1}\widehat{\tau}_{0}^2 \right)^2}{B_{BOOT}(B_{CV}-1)} \right).$$
Thus fixing $B_{BOOT}\times B_{CV}=N_T$, the variance is minimized when 
$$B_{BOOT}\approx \frac{\widehat{\sigma}^2_{BT}}{\widehat{\tau}^2_{0}}\times N_T~~~\mbox{ and }~~~ B_{CV}\approx \frac{\widehat{\tau}^2_{0}}{\widehat{\sigma}^2_{BT}}.$$
It suggests that the optimal number of cross-validation per bootstrap should be approximately constant, whose value 
shouldn't change with the budget for the total number of model training. Normally, $\widehat{\tau}^2_{0}$ can be substantially greater than $\widehat{\sigma}_{BT}^2$ and $B_{CV}$ should be set to be close to their ratio. In the toy example, this ratio is approximately 20. 
\end{remark}

\begin{remark}
The number of distinct observations used in training the bootstrapped prediction model is smaller than $m_{adj}$. Specifically, the number of distinct observations in bootstrapped training set is on average only $0.632\times m_{adj}.$ Therefore, there is a tendency that the ``effective total sample size" in bootstrap procedure is smaller than $n,$ which may cause an upward bias in estimating the variance of $\widehat{Err}_m^{CV}$ using the bootstrap variance estimator $\widehat{\sigma}_m^2.$ To correct this bias, we can consider deflating the bootstrap variance estimate by  a factor $(n-m_{adj}+0.632m_{adj})/n=1-0.368m_{adj}/n,$ i.e., $\left(\widehat{\sigma}_{m, adj}^{CV}\right)^2=\left(\widehat{\sigma}_{m}^{CV}\right)^2(1-0.368m_{adj}/n),$
which accounts for the reduction of effective sample size in bootstrapped training set. 
\end{remark}

\begin{remark}
The proposed method focuses on data with i.i.d. observations, a standard setting for both cross-validation and bootstrap. On the other hand,  it is possible to modify our algorithm to take into account correlations among observations in specific applications. For example, while the longitudinal outcomes from the same subject were correlated, data from different subjects can still be treated as independent observations. Thus, the proposed method can be used by cross-validating and bootstrapping independent subjects. Observations from a time series often demonstrate strong serial correlations. To account for this type of correlation, one may break the time series into smaller blocks consisting of successive observations and cross-validating and bootstrapping the resulting blocks. With an appropriately chosen block size, the correlation between blocks becomes ignorable in comparison with the within-block correlation, which is maintained by using the block as a unit in cross-validation and bootstrap.
\end{remark}

Sometimes, training the prediction model can be very expensive in terms of computation, and it may not be feasible to conduct even the accelerated bootstrap in algorithm \ref{ag:mainboot}. 
In such a case, regardless of the selection of $B_{BOOT}$ and $B_{CV}$, the Monte Carlo error in estimating the bootstrap variance may not be ignorable. Consequentially, 
$$\frac{\sqrt{n}(\widehat{Err}_m^{CV}-Err_m)}{\widehat{\sigma}_m^{CV}}~~~\mbox{ or }~~~ \frac{\sqrt{n}(\widehat{Err}_m^{CV}-Err_m)}{\widehat{\sigma}_{m, adj}^{CV}}$$
may not follow $N(0, 1).$ If we can approximate this distribution, then one can still construct a 95\% CI for $Err_m$ based on $(\widehat{Err}_m^{CV}, \widehat{\sigma}_m^{CV}).$ One analogy is that the CI for the mean of a normal distribution can be constructed using the t distribution rather than the normal distribution. 
Let $\sigma_m^{CV}(\infty)$ be the bootstrap variance estimator if both $B_{BOOT}$ and $B_{CV} \rightarrow \infty,$

\begin{equation}\frac{\sqrt{n}(\widehat{Err}_m^{CV}-Err_m)}{\widehat{\sigma}_m^{CV}}=\frac{\sqrt{n}(\widehat{Err}_m^{CV}-Err_m)}{\sigma_m^{CV}(\infty)}\times \frac{\sigma_m^{CV}(\infty)}{\widehat{\sigma}_m^{CV}}. \label{eq:zscoredecomp}\end{equation}
The first term of the left-hand side of (\ref{eq:zscoredecomp}) should be approximated well by a standard Gaussian distribution since the ``ideal" bootstrap variance estimator is used. The second term is independent of the first term and reflects the Monte-Carlo variation of approximating $\sigma_m^{CV}(\infty)$ via finite numbers of bootstrap and cross-validation iterations. To approximate the distribution of this ratio, we can additionally bootstrap the variance estimator. 
This observation motivated the inclusion of following steps (algorithm \ref{ag:calboot}) after lines (\ref{line1:bootest}-\ref{line2:bootest}) of algorithm \ref{ag:mainboot}, when very small $B_{BOOT}$ and $B_{CV}$ are used.

\begin{algorithm}[ht]
\caption{Bootstrap Calibration}\label{ag:calboot}
{\small
\begin{algorithmic}[1]
\setcounter{ALG@line}{12}
\For{ $l\gets 1$ to  $L$} 
\State Construct the bootstrapped dataset
$\Theta_l^*=\left\{\theta^{**}_{lbk}, b=1,\cdots, B_{BOOT}; k=1, \cdots, B_{CV}\right\},$
where the vector  $(\theta^{**}_{lb1}, \cdots, \theta^{**}_{lbK})$ is a random sample from $B_{BOOT}$ vectors 
$\{(\theta^*_{b1}, \cdots, \theta^*_{bK}), b=1, \cdots, B_{BOOT}\}.$
\State Let
$$\widehat{\sigma}_{l,BT}^{2*}=\frac{1}{B_{BOOT}-1}\sum_{b=1}^{B_{BOOT}} (\bar{\theta}^{**}_{lb}-\bar{\theta}^{**}_l)^2-\frac{1}{B_{CV}(B_{CV}-1)B_{BOOT}}\sum_{b=1}^{B_{BOOT}} \sum_{k=1}^{B_{CV}}(\theta^{**}_{lbk}-\bar{\theta}^{**}_{lb})^2,$$
where 
$\bar{\theta}_{lb}^{**}=B_{CV}^{-1}\sum_{k=1}^{B_{CV}} \theta^{**}_{lbk},~~\mbox{and}~~\bar{\theta}^{**}_l=B_{BOOT}^{-1}\sum_{b=1}^{B_{BOOT}} \bar{\theta}^{**}_{lb}.$
\EndFor
\State Let 
$$Z_l^*=Z_l\frac{\widehat{\sigma}_{BT}}{\widehat{\sigma}_{l, BT}^{*}} , l=1, \cdots, L,$$
where $Z_l\sim N(0, 1),$ and use the distribution of $Z_l^*$ to approximating that of  $$\frac{\sqrt{n}(\widehat{Err}^{CV}_m-Err_m)}{\sigma_m^{CV}(\infty)} \times \frac{\sigma_{m}^{CV}(\infty)}{\widehat{\sigma}_m^{CV}}.$$
\State Find the cut off value $c_{1-\alpha/2}$ such that 
$L^{-1}\sum_{l=1}^L I\left(|Z_l^*|<c_{1-\alpha/2} \right)=1-\alpha.$
\State The final $(1-\alpha)100$\% CI for $Err_m$ is 
$$\left[\widehat{Err}^{CV}_m-c_{1-\alpha/2}\times \widehat{\sigma}_m^{CV},  \widehat{Err}^{CV}_m+c_{1-\alpha/2}\times \widehat{\sigma}_m^{CV} \right].$$
\end{algorithmic}
}
\end{algorithm}

This resulting CI is expected to be wider than that generated by the algorithm \ref{ag:mainboot}.  This is a necessary cost to pay for using small numbers of bootstrap and cross-validation iterations. Note that although two bootstraps have been used in the modified algorithm, the increase in computational burden is minimal,  since these two steps are not nested but sequential. The practical time savings are examined by the simulation study reported in Section 5 of the Supplementary Material \citep{cai2025bootstrapping}. We have found that algorithm \ref{ag:calboot} can be substantially faster than algorithm \ref{ag:mainboot}, when the model training is complicated and time demanding. 
The performance of this method depends on the normal approximation to the distribution of $$(\widehat{Err}_m^{CV}-Err_m)/\sigma_m^{CV}(\infty)$$ 
and the bootstrap approximation to the distribution of 
$$\sigma_m^{CV}(\infty)/\widehat{\sigma}_m^{CV}.$$  The second bootstrap is a calibration step for producing a CI of $Err_m$ with a coverage level comparable to that based on $(\widehat{Err}_m^{CV}-Err_m)/\sigma_m^{CV}(\infty)$. If the latter yields a CI with poor coverage, then the new CI would suffer the same limitation. Operationally, one may choose a slightly bigger $B_{CV}$, e.g., $(B_{BOOT}, B_{CV})=(20, 50)$ 
to avoid a negative variance component estimator in the random effects model.

\section{Application 1}

\subsection{Theoretical Properties of Application 1 (Precision Medicine)}
In the first application, we are interested in evaluating the performance of a precision medicine strategy. In this case, it is not difficult to verify conditions C1-C5 in Section 1 of the Supplementary Material \citep{cai2025bootstrapping} under suitable assumptions. For example, if the matrix $A_0=E(\tilde{Z}\tilde{Z}')$ is nonsingular, $\Pr(G=1)=\pi\in (0, 1)$, and $G\perp Z$, i.e., the treatment assignment is randomized,  then $\widehat{\gamma}$ and $\widehat{\beta}$ converge to deterministic limits $\gamma_0$ and $\beta_0,$ respectively, as $n \rightarrow \infty$, and especially
$$ \sqrt{n}(\widehat{\beta}-\beta_0)=\frac{1}{\sqrt{n}}\sum_{i=1}^n \left\{\pi(1-\pi)A_0\right\}^{-1}\left\{Y_i-\gamma_0'\tilde{Z}_i-(G_i-\pi)\beta_0'\tilde{Z}_i\right\}+o_p(1),$$
where $\beta_0$ is an unique minimizer of $ m(\beta)=\E\left\{\left(Y^{(1)}-Y^{(0)}-\beta'\tilde{Z}\right)^2\right\}.$ Thus condition C1 is satisfied.  Second,  the classes of functions $\{y\I(\beta'z>0) \mid \beta \in \Omega\}$ , $\{y\I(\beta'z\le 0) \mid \beta \in \Omega\}$, $\{\I(\beta'z>0)\mid \beta\in \Omega\}$, and $\{\I(\beta'z\le 0)\mid \beta\in \Omega\}$ are Donsker, where $\Omega$ is a compact set in $R^{p+1}$.  This fact suggests that the stochastic processes
{\small 
\begin{align*}
&\frac{1}{\sqrt{n}}\sum_{i=1} \left[Y_i\I(\beta'\tilde{Z}_i>0)-\E\left\{Y\I(\beta'\tilde{Z}>0)\right\} \right],~~\frac{1}{\sqrt{n}}\sum_{i=1} \left[Y_i\I(\beta'\tilde{Z}_i\le 0)-\E\left\{Y\I(\beta'\tilde{Z}\le 0)\right\} \right]\\
&\frac{1}{\sqrt{n}}\sum_{i=1} \left[\I(\beta'\tilde{Z}_i>0)-\Pr(\beta'\tilde{Z}>0) \right], ~~\frac{1}{\sqrt{n}}\sum_{i=1} \left[\I(\beta'\tilde{Z}_i\le 0)-\Pr(\beta'\tilde{Z}\le 0) \right]
\end{align*}
}
are all stochastically continuous in $\beta.$ Therefore, the processes
{\small 
$$U_1(\beta)=\sqrt{n}\left\{\frac{\sum_{i=1}^n Y_iG_i\I(\beta'\tilde{Z}_i>0)}{\sum_{i=1}^n G_i \I(\beta'\tilde{Z}_i>0)}-\frac{\sum_{i=1}^n Y_i(1-G_i)\I(\tilde{\beta}'Z_i>0)}{\sum_{i=1}^n (1-G_i)\I(\tilde{\beta}'Z_i>0)}-\E\left(Y^{(1)}-Y^{(0)} \mid \beta'\tilde{Z}>0 \right)\right\}$$
}
and its counterpart $U_0(\beta)$ are also stochastically continuous in $\beta$, i.e., $U_g(\beta_1)-U_g(\beta_2)=o_p(1)$ for $\|\beta_2-\beta_1\|=o(1), g\in \{0, 1\}.$ As a consequence, condition C2 is satisfied. It is clear that
$$l_1(\beta)=\E\left(Y^{(1)}-Y^{(0)} \mid \beta'\tilde{Z}>0 \right)~~~\mbox{ and }~~~ l_0(\beta)=\E\left(Y^{(1)}-Y^{(0)} \mid \beta' \tilde{Z}\le 0 \right)$$
are differentiable in $\beta$ in a small neighborhood of $\beta_0,$ if the random variable $\beta_0'\tilde{Z}$ has a continuously differentiable bounded density function and $\E\left(Y^{(g)}\mid \beta_0'\tilde{Z}=s\right)$ is smooth in $s$.  This suffices for condition C3.  Next, the central limit theorem and the delta method together imply that $U_g(\beta_0)$ converges weakly to a mean zero Gaussian distribution as $n \rightarrow \infty,$ where $g\in \{0, 1\}.$ Lastly, 
\begin{align*}
&|\E\{\Pr(Y^{(1)}-Y^{(0)}\mid \widehat{\beta}'\tilde{Z}>0)\}-\Pr(Y^{(1)}-Y^{(0)}\mid \beta_0'\tilde{Z}>0)| + \\
&|\E\{\Pr(Y^{(1)}-Y^{(0)}\mid \widehat{\beta}'\tilde{Z}\le 0)\}-\Pr(Y^{(1)}-Y^{(0)}\mid \beta_0'\tilde{Z}\le 0)| =o_p(|\E(\widehat{\beta})-\beta_0|)=o_p(n^{-1/2})
\end{align*}
and C5 is satisfied. 


\subsection{Simulation Study}
In the simulation study, we considered two settings corresponding to low and high dimensional covariates vector $Z_i.$ The covariate $Z_i$ was generated from a $p$-dimensional standard multivariate Gaussian distribution and the continuous outcome $Y_i^{(g)} $ was generated via two linear regression models:
$$ Y_i^{(g)}=\beta_g'\tilde{Z}_i+\epsilon_i^{(g)}, g\in \{0, 1\},$$

\noindent{}where $\beta_g=(0, 0.25, (-1)^{g+1}0.25, 0.25, (-1)^{g+1}0.25, 0, \cdots, 0)'$, 
and $\epsilon_i^{(g)}\sim N(0, 1), g\in \{0, 1\}.$ The treatment assignment indicator $\{G_1, \cdots, G_n\}$ was a random permutation of $\{1, \cdots, 0, \cdots\}$ consisting of half ones and half zeros. The observed outcome $Y_i=Y_i^{(1)}G_i+Y_i^{(0)}(1-G_i).$  The generated data $D_n=\left\{(Y_i, G_i, Z_i), i=1, \cdots, n\right\}.$

In the first set of simulations, we let $p=10$ and the sample size $n=180$.  We considered the cross-validation estimator $\widehat{Err}_m^{CV}$ for $Err_m, m\in\{80, 90, 100, 110, 120, 130, 140\}.$ Due to symmetry, we only considered the case where $Err_m$ was the ATE among patients recommended to receive treatment $G=1,$ i.e., responders. The true treatment effect among responders was calculated with an independently generated test set consisting of 200,000 patients. The true $Err_m$ is 0.37, 0.39, 0.40, 0.42, 0.43, and 0.44 for $m=$ 80, 90, 100, 110, 120, 130, and 140, respectively. The increasing trend in $Err_m$ reflected the improved quality of the estimated ITR score based on a larger training set. 
The ATE among the responders based on true individualized treatment effects was 0.56.

We constructed the cross-validation estimate of $Err_m$ from 1,000 datasets $D_n$.  For each simulated dataset $D_n$, we divided the dataset into a training set of size $m$ and a testing set of size $n-m$. The ITR score $\widehat{\Delta}(z\mid D_{train})$ was estimated based on the training set and responders in the testing set were identified. The ATE estimator among responders in the test set was simply the mean difference in $Y$ between responders who received the active treatment $G=1,$ and responders who received the control treatment $G=0.$ This process was repeated 400 times and the resulting $\widehat{Err}_m^{CV}$ was the sample average of 400 ATE estimators among identified responders in the testing set.  In addition, we used the proposed bootstrap method to compute the standard error estimators $\widehat{\sigma}_m^{CV}$ and $\widehat{\sigma}_{m,adj}^{CV}$ 
with $(B_{BOOT}, B_{CV})=(400, 20).$ 
The 95\% CI for $Err_m$ was also constructed for each simulated dataset.  We also examined the performance of constructed CI using only a very small number of bootstrap iterations, choosing $(B_{BOOT}, B_{CV})=(20, 25).$  

{In the second set of simulations with $p=1000,$ the estimated regression coefficient $\widehat{\beta}$ in the ITR score $\widehat{\Delta}(z\mid D_{train})$ was estimated via lasso regularization. Specifically, we estimated $\beta$ by minimizing a regularized loss function
 $$\sum_{X_i\in D_{train}} \left[Y_i-\gamma'\tilde{Z}_i-(G_i-\pi)\beta'\tilde{Z}_i\right]^2 +\lambda_1|\gamma_Z|_1+\lambda_2|\beta_Z|_1,$$
 where $\lambda_1$ and $\lambda_2$ were appropriate penalty parameters and $\gamma_Z$ and $\beta_Z$ are vectors of $\gamma$ and $\beta$ excluding the first component, respectively. To save computational time, both penalty parameters were fixed at 0.10 in all simulations instead of being adaptively selected via cross-validation within the training set.  
 Similar to the low-dimensional case, we simulated 1,000 datasets and for each generated dataset $D_n,$ we calculated $Err(D_n),$ $\widehat{Err}_m^{CV}$, the bootstrap standard error estimators $\widehat{\sigma}_m^{CV}$ and $\widehat{\sigma}_{m,adj}^{CV}$, and the corresponding 95\% CIs.  We also investigated the performance of the CIs constructed using different number of bootstraps and cross validations.

 The simulation results including the true value of $Err_m$, the empirical average and standard deviation of the cross-validation estimator $\widehat{Err}_m^{CV}$, and the empirical coverage of 95\% CIs based on $B_{BOOT}=400$ are summarized in Table \ref{tab:simupm}.  In addition, the empirical coverage levels of the constructed 95\% CIs based on $B_{BOOT}=20$ were summarized in Table \ref{tab:simupmsmall}.
 For both low- and high-dimensional cases, the cross-validated estimator $\widehat{Err}^{CV}_m$ was almost unbiased in estimating $Err_m$, especially relative to the empirical standard deviation of $\widehat{Err}^{CV}_m$. The empirical coverage level of 1000 constructed 95\% CIs for $Err_m$ based on the bootstrap variance estimator from a large number of bootstrap iterations was quite close to its nominal level. After sample size adjustment in variance estimation, the constructed CIs based on $\widehat{\sigma}_{m,adj}^{CV}$ slightly under-covered the true parameter with empirical coverage levels between 90\% and 93\%.  When using a small number of bootstrap iterations ($B_{BOOT}=20$), the proposed bootstrap calibration can be used to maintain a proper coverage level (Table \ref{tab:simupmsmall}). As a price, the median width of the calibrated CI increased 14-28\%. Note that the theoretical justification for the Gaussian approximation to the cross-validated estimator in the high-dimensional case was not provided. However, the empirical distribution of $\widehat{Err}^{CV}_m$ was quite ``Gaussian" with its variance being approximated well by the bootstrap method. This observation ensured the good performance of resulting 95\% CIs. In addition, the empirical coverage levels of the 95\% CIs of $Err_{m}$ with respect to $Err(D_n)$ were 92.9\% and 95.4\% in low- and high-dimensional settings, respectively, where $(m, n)=(140, 180).$

\begin{table}[h]
\setlength{\tabcolsep}{5pt}
\caption{Simulation results for precision medicine. $Err_m$, the true ATE in identified high value subgroup;  $E(\widehat{Err}_m^{CV}),$ the empirical average of the cross-validation estimate $\widehat{Err}_m^{CV}$; SD, the empirical standard deviation of the cross-validation estimate $\widehat{Err}_m^{CV}$  ; COV-adj, the empirical coverage level of 95\% CIs based on $\widehat{\sigma}_{m, adj}^{CV}$ from bootstrap; COV, the empirical coverage level of 95\% CIs based on $\widehat{\sigma}_{m}^{CV}$ from bootstrap.}
\begin{tabular}{ c ccccc c ccccc }
\hline
$m$ & \multicolumn{5}{c}{$p = 10$} & & \multicolumn{5}{c}{$p = 1000$} \\
 & $Err_m$ & $\E(\widehat{Err}_m^{CV})$ & SD   & $\Cov$-adj & $\Cov$ & & $Err_m$ & $\E(\widehat{Err}_m^{CV})$ & SD   & $\Cov$-adj & $\Cov$ \\
 \cline{2-6} \cline{8-12}\\
80  & 0.369 & 0.377 & 0.196 & 92.7\% & 95.1\% && 0.079 & 0.082 & 0.190& 92.7\% & 95.9\%\\
90  & 0.384 & 0.395 & 0.197 & 91.9\% & 95.2\% && 0.094 & 0.095 & 0.198& 92.0\% & 96.1\%\\
100 & 0.398 & 0.409 & 0.198 & 92.3\% & 95.1\% && 0.108 & 0.110 & 0.208& 91.1\% & 95.9\% \\
110 & 0.412 & 0.422 & 0.199 & 91.9\% & 95.3\% && 0.119 & 0.123 & 0.218& 91.3\% & 95.6\% \\
120 & 0.421 & 0.433 & 0.199 & 91.4\% & 95.1\% && 0.135 & 0.138 & 0.229& 91.0\% & 95.1\& \\
130 & 0.431 & 0.441 & 0.201 & 90.7\% & 95.2\% && 0.152 & 0.152 & 0.242& 90.3\% & 95.0\% \\
140 & 0.439 & 0.449 & 0.202 & 91.4\% & 95.4\% && 0.166 & 0.166 & 0.256& 89.4\% & 94.9\% \\
\end{tabular}
\label{tab:simupm}
\end{table}

\begin{table}[h]
\setlength{\tabcolsep}{5pt}
\caption{Empirical coverage levels of 95\% CIs of the ATE among responders using various numbers of bootstrap iterations.}
\begin{tabular}{ cccccccc }
\hline\\
$m$   & \multicolumn{3}{c}{CIs Based on $\widehat{\sigma}_{m,adj}^{CV}$} && \multicolumn{3}{c}{CIs based on $\widehat{\sigma}_m^{CV}$}\\
& Algorithm \ref{ag:mainboot} & Algorithm \ref{ag:mainboot} & Algorithm \ref{ag:calboot} && Algorithm \ref{ag:mainboot} & Algorithm \ref{ag:mainboot} & Algorithm \ref{ag:calboot}\\ 
 & $B_{BOOT}=400$  & $B_{BOOT}=20$  & $B_{BOOT}=20$  && $B_{BOOT}=400$  & $B_{BOOT}=20$ & $B_{BOOT}=20$\\
     \cline{2-4} \cline{6-8}\\
      & \multicolumn{7}{c}{$p = 10$} \\
      \cline{2-8}\\
80   & 92.7\% & 90.8\% & 94.8\% && 95.1\% & 93.7\% & 96.6\%\\
90   & 91.9\% & 90.2\% & 94.1\% && 95.2\% & 92.7\% & 96.8\%\\
100  & 92.3\% & 90.3\% & 94.6\% && 95.1\% & 93.0\% & 97.4\%\\
110  & 91.9\% & 89.8\% & 95.3\% && 95.3\% & 93.5\% & 97.6\%\\
120  & 91.4\% & 89.0\% & 96.7\% && 95.1\% & 93.3\% & 98.6\%\\
130  & 90.7\% & 89.2\% & 96.5\% && 95.2\% & 92.4\% & 98.7\%\\
140  & 91.4\% & 86.2\% & 97.1\% && 95.4\% & 90.9\% & 98.3\%\\
& \multicolumn{7}{c}{$p = 1000$} \\
\cline{2-8}\\
80  &  92.7\% & 89.9\% & 94.2\% && 95.9\% & 93.1\% & 96.7\%\\
90  &  92.0\% & 90.7\% & 95.4\% && 96.1\% & 93.9\% & 97.4\%\\
100 &  91.1\% & 89.6\% & 95.8\% && 95.9\% & 93.9\% & 97.9\%\\
110 &  91.3\% & 89.0\% & 95.0\% && 95.6\% & 93.7\% & 97.3\%\\
120 &  91.0\% & 89.0\% & 94.5\% && 95.1\% & 93.8\% & 97.9\%\\
130 &  90.3\% & 87.5\% & 95.0\% && 95.0\% & 93.6\% & 97.7\%\\
140 &  89.4\% & 86.5\% & 96.0\% && 94.9\% & 92.6\% & 98.1\%
\end{tabular}
\label{tab:simupmsmall}
\end{table} 
 
In summary, the proposed CIs based on the bootstrap standard error estimator $\widehat{\sigma}_m^{CV}$ have a good coverage level. The bootstrap calibration effectively corrects the under-coverage of the CIs based on a very small number of bootstraps. The constructed CI can be viewed as a CI for both $Err(D_n)$ and $Err_m,$ when $m\approx n$. In this case, due to the complexity of the evaluation procedure in the test set, no existing method is readily available to study the distribution of the cross-validation estimator for the ATE among ``responders''. }
 
\subsection{Real Data Example}
The PEACE trial was designed to examine the effect of ACEi on reducing future cardiovascular events in patients with stable coronary artery disease and normal or slightly reduced left ventricular function \citep{peace2004angiotensin}. While a total of 8290 patients are enrolled in the study, we focus on a subgroup of 7865 patients with complete covariate information, in which 3947 and 3603 patients were assigned to receive ACEi and placebo, respectively.  The endpoint of the study is survival time and the estimated hazard ratio was 0.92 (95\% CI: 0.78 to 1.08) with a non-significant p-value of 0.30. 
The objective of our analysis was to identify a high value subgroup of patients who may benefit from ACEi, even though the ATE of ACEi in the entire study population was not significant. 
To build a candidate scoring system capturing the ITR, we used 4 baseline covariates previously identified as statistically and clinically important predictors of overall mortality \citep{solomon2006renal}: age, gender, eGFR for renal function, and left ventricular ejection fraction. Separate Cox proportional hazards models were fitted in the ACEi and placebo arms, and the between-group difference in estimated RMST was used to compute the ITR score. The ITR score was derived from a training set of size $m=6292$ (80\% of the study population).  

The ATE in the test set can be measured in different ways. We first considered the ATE as the RMST difference. Based on 500 cross-validations, the cross-validated estimate $\widehat{Err}^{CV}_m$ was 21.1 days for $\widehat{\Delta}_1(D_{train}, D_{test}),$ the ATE in the high-value subgroup, and -13.2 days for $\widehat{\Delta}_0(D_{train}, D_{test}),$ the ATE in the complement of the high-value subgroup. Their difference, i.e., the interaction with treatment, was 34.3 days.  Now, we implement the proposed method to estimate the standard errors of these estimators. Specifically, we let $(B_{BOOT}, B_{CV})=(500,20)$. The 95\% CI of the ATE in the high-value subgroup was [-1.3, 45.5] days ($p=0.064$). 
The 95\% CI of the ATE in the complement of the high-value subgroup was [-31.5, 5.2] days ($p=0.161$). 
The 95\% CI of their difference was [4.3, 64.3] days ($p=0.025$), 
suggesting that the ITR score constructed from the training set of 6,292 patients had a statistically significant interaction with the treatment, i.e. the ATE in the high-value subgroup was higher than in the remaining patients. The detailed results are summarized in Table \ref{tab:peace}, which also reports the results on the hazard ratio.

\begin{table}[ht]
\centering
\caption{Results of estimated the ATE among identified responders in PEACE trial}
\begin{tabular}{ crrc c cccc }
\hline
    \multicolumn{4}{c}{restricted mean survival time} & & \multicolumn{4}{c}{hazard ratio} \\
 $Err_m$  &  $\widehat{Err}_m^{CV}$ & 95\% CI for $Err_m$   & $p$-value & & $Err_m$ & $\widehat{Err}_m^{CV}$ & 95\% CI for $Err_m$   & $p$-value\\
    \cline{1-4} \cline{6-9}\\
$\widehat{\Delta}_1$ & 21.1 & [-1.3,43.5] & 0.064 && $\widehat{\Delta}_1$ & 0.80 & [0.66, 0.98]& 0.028 \\ 
$\widehat{\Delta}_0$   & -13.2 & [-31.5, 5.2]& 0.161 && $\widehat{\Delta}_0$ & 1.18 & [0.85, 1.65]& 0.326 \\
$\widehat{\Delta}_1-\widehat{\Delta}_0$ & 34.3 & [4.3, 64.3] & 0.025 && $\widehat{\Delta}_1/\widehat{\Delta}_0$ & 0.68 & [0.44, 1.04] &  0.076  \\
\end{tabular}
\label{tab:peace}
\end{table}

\section{Application 2}

\subsection{Theoretical Properties of Application 2 (Binary Outcomes)}
In the second application, we are interested in estimating the c-index from a logistic regression model via cross-validation. In this case, it is not difficult to verify the conditions C1-C5 in Section 1 of the Supplementary Material \citep{cai2025bootstrapping} under conventional assumptions.  For example, under the condition that there is no $\beta$ such that the hyperplane $\beta'z=a_0$ can perfectly separate observations with $Y_i=1$ from those with $Y_i=0$ and the matrix 
$$A_0=\E\left[\tilde{Z}'\tilde{Z} \frac{\exp(\beta'\tilde{Z})}{\left\{1+\exp(\beta'\tilde{Z})\right\}^2}  \right]$$
is positive definite for all $\beta$, the maximum likelihood estimator based on the logistic regression, $\widehat{\beta},$ converges to a deterministic limit $\beta_0$ in probability as $n \rightarrow \infty$ and
$$\sqrt{n}(\widehat{\beta}-\beta_0)=\frac{1}{\sqrt{n}}\sum_{i=1}^n A_0^{-1}\left(Y_i-\frac{\exp(\beta_0'\tilde{Z}_i)}{1+\exp(\beta_0'\tilde{Z}_i)}\right)+o_p(1),$$
and thus the condition C1 is satisfied \citep{tian2007model}.
Second, the class of functions $\{\I(\beta'\tilde{z}<0)\mid \beta\in \Omega\}$  is Donsker, where $\Omega$ is a compact set in $R^{p+1}$.  This fact suggests that the U-process
$$U(\beta)=\sqrt{n}\left[L\left(D, \beta\right)-\Pr\left(\beta'(\tilde{Z}_1-\tilde{Z}_2)<0\mid Y_1=0, Y_2=1\right)\right]$$
is stochastically continuous, where 
$$L\left(D, \beta\right)=\frac{1}{m_0m_1}\sum_{Y_i=0}\sum_{Y_j=1}\I(\beta'\tilde{Z}_i<\beta'\tilde{Z}_j),
$$ 
$m_g=\sum_{i=1}^n \I(Y_i=g).$ As a consequence, condition C2 is satisfied as $n \rightarrow \infty$ and $0<\Pr(Y=1)<1.$ 
It is clear that $\Pr\left(\beta'(\tilde{Z}_1-\tilde{Z}_2)<0\mid Y_1=0, Y_2=1\right)$ is differentiable in $\beta$ in a small neighborhood of $\beta_0,$ if $\beta_0'\tilde{Z}$ has a differentiable density function, which suffices for condition C3.  Next, the central limit theorem for U-statistics implies that
$$\sqrt{n}\left[L\left(D, \beta_0\right)-\Pr\left(\beta_0'(\tilde{Z}_1-\tilde{Z}_2)<0\mid Y_1=0, Y_2=1\right)\right]$$
converges weakly to a mean zero Gaussian distribution as $n \rightarrow \infty.$ Lastly, 
\begin{align*}
&\E\left\{\Pr\left(\widehat{\beta}'(\tilde{Z}_1-\tilde{Z}_2)<0\mid Y_1=0, Y_2=1\right)\right\}-\Pr\left(\beta_0'(\tilde{Z}_1-\tilde{Z}_2)<0\mid Y_1=0, Y_2=1\right) \\
=&O(|\E(\widehat{\beta})-\beta_0|)=o_p(n^{-1/2})
\end{align*}
and C5 is satisfied. Therefore, we expect that $\widehat{Err}_m^{CV}-Err_m$ is approximately Gaussian whose variance can be consistently estimated using the proposed bootstrap method. Note that we do not assume that the logistic regression model is correctly specified. 

\subsection{Simulation Study}
In the numerical study, $Z_i$ followed a $p$-dimensional standard multivariate normal distribution and the binary outcome $Y_i$ followed a Bernoulli distribution 
$$\Pr(Y_i=1|Z_i)=\mbox{expit}(\beta_0'\tilde{Z}_i),$$
where $\beta_0=(0, 1.16, 1.16, 1.16, 1.16, 0, \cdots, 0)'.$ This regression coefficient was selected such that the misclassification error of the optimal Bayesian classification rule was approximately 20\%. In the first set of simulations, we let $(p, n)=(10, 90)$  and considered the cross-validation estimator $\widehat{Err}_m^{CV}$ for the c-index $Err_m, m\in\{40, 45, 50, 55, 60, 65, 70, 75, 80\}$. Based on results from 1,000 datasets, we summarized the empirical average and standard deviation of $\widehat{Err}_m^{CV}$ for c-index and the empirical coverage level of 95\% CIs based on bootstrap variance estimates. We also examined the performance of the bootstrap calibration in algorithm \ref{ag:calboot} for constructing 95\% CIs with a very small number of bootstraps. To this end, we set $(B_{BOOT}, B_{CV})=(20, 50)$. In the second set of simulations with $p=1000$, the regression coefficient was estimated with lasso regularization. The simulation results were reported in Section 4 of the Supplementary Material \citep{cai2025bootstrapping}. The cross-validation estimator $\widehat{Err}_m^{CV}$ was almost unbiased in estimating $Err_m$ with its empirical bias negligible in comparison with the standard deviation of $\widehat{Err}_m^{CV}.$ The empirical coverage level of 95\% CIs with $B_{BOOT}=400$ was fairly close to the nominal level. When a small number of bootstraps were used, the coverage level of the CIs was markedly lower than those constructed via a large number of bootstraps.  However, with the proposed bootstrap calibration in algorithm \ref{ag:calboot}, the coverage level became comparable to those using more bootstrap replicates. 

\subsection{Real Data Examples}
In this example, we tested our proposed method on the MI dataset from the UCI machine learning repository. The dataset contained 1700 patients and up to 111 predictors collected from hospital admission up to 3 days after admission for each patient. We were interested in predicting all-cause mortality. After removing features with more than 300 missing values, there were 100 prediction features available at day 3 after admission, including 91 features available on admission. The observed data consisted of 652 patients with complete information on all 100 features.  There were 72 binary, 21 ordinal, and 7 continuous features. Of 652 patients, there were 62 deaths, corresponding to a cumulative mortality of 9.5\%. We considered training size $m\in \{196, 261, 326, 391, 456, 522, 587\},$ which represented 30\% to 90\% of the total sample size.  We considered four prediction models, all trained by fitting a lasso regularized logistic regression.  Model 1 was based on 91 features collected at the time of admission; Model 2 was based on 100 features collected up to day 3 after hospital admission; Model 3 was based on 126 features collected at the time of admission after converting all ordinal features into binary features;  and Model 4 was based on 159 features collected up to day 3 after converting all ordinal features into binary features.

First, we estimated the cross-validated c-index 
based on 500 random cross-validations. We then constructed the 95\% CI based on the proposed bootstrap method with $(B_{BOOT}, B_{CV})=(400, 20).$ The results were reported in Table \ref{tab:example2a1}.  Model 2, which included 8 additional features, had a slightly better prediction performance than Model 1. 
Likewise, Model 3 and Model 4 were inferior to Models 1 and 2, respectively, suggesting that converting ordinal predictive features into multiple binary features may have a negative impact. 
We then formally compared the performance of the models by constructing the 95\% CI for the difference in c-index between Models 1 and 2; Models 1 and 3; and Models 2 and 4. The detailed results of the comparisons are reported in Table \ref{tab:example2a2}.  All CIs include zero, suggesting that none of the observed differences in c-index is statistically significant at the 0.05 level.

In the second example of a binary outcome, we tested our proposal on the “red wine” dataset studied. 
The results were reported in Section 3 of the Supplementary Material \citep{cai2025bootstrapping}, where the proposed method was used to make inferences on the ROC curve itself, which can be estimated by repeating $K$-fold cross-validation multiple times.

\begin{table}[ht]
\centering
\setlength{\tabcolsep}{5pt}
\caption{Results of estimating the c-index based on MI data from UCI Repository.}
\begin{tabular}{ ccc c cc c cc c cc}
\hline
$m$ & AUC & 95\% CI && AUC & 95\% CI && AUC & 95\% CI && AUC & 95\% CI  \\
& \multicolumn{2}{c}{91 features, Day 0} && \multicolumn{2}{c}{100 features, Day 3} && \multicolumn{2}{c}{126 features, Day 0} && \multicolumn{2}{c}{159 features, Day 3} \\
\hline\\
196 &  0.711  & [0.645, 0.778] && 0.711 & [0.643, 0.779] &&  0.676  & [0.596, 0.755] && 0.664 & [0.583, 0.745] \\
261 &  0.729  & [0.660, 0.798] && 0.731 & [0.659, 0.802] &&  0.692  & [0.610, 0.773] && 0.678 & [0.594, 0.762] \\
326 &  0.743  & [0.672, 0.814] && 0.747 & [0.674, 0.820] &&  0.700  & [0.616, 0.784] && 0.688 & [0.600, 0.776] \\
391 &  0.753  & [0.681, 0.825] && 0.760 & [0.687, 0.833] &&  0.712  & [0.628, 0.796] && 0.702 & [0.614, 0.791] \\
456 &  0.759  & [0.688, 0.831] && 0.768 & [0.695, 0.841] &&  0.716  & [0.631, 0.801] && 0.709 & [0.620, 0.798] \\
522 &  0.766  & [0.694, 0.837] && 0.777 & [0.705, 0.850] &&  0.723  & [0.640, 0.806] && 0.718 & [0.630, 0.805] \\
587 &  0.771  & [0.702, 0.840] && 0.785 & [0.718, 0.852] &&  0.729  & [0.644, 0.814] && 0.727 & [0.635, 0.819] \\
\end{tabular}
\label{tab:example2a1}
\end{table}

\begin{table}[ht]
\centering
\setlength{\tabcolsep}{5pt}
\caption{Results of comparing c-index between different prediction models based on MI data from UCI repository; Model 1: 91 features at Day 0; Model 2: 100 features at Day 3; Model 3: 126 features at Day 0; Model 4: 159 features at Day 3 }
\begin{tabular}{ ccc c cc c cc }
\hline
$m$   & $\Delta$ AUC  & 95\% CI &&  $\Delta$ AUC & 95\% CI && $\Delta$ AUC & 95\% CI\\
    & \multicolumn{2}{c}{ Models 2 vs. 1 $(\times 10^{-2})$} && \multicolumn{2}{c}{Models 1 vs. 3 $(\times 10^{-2})$} && \multicolumn{2}{c}{Models 2 vs. 4 $(\times 10^{-2})$}  \\
    \cline{2-3} \cline{5-6} \cline{8-9}\\
196 & -0.042 & [-2.361, 2.276] && 3.562 & [-1.400, 8.524] && 4.714 & [-0.647, 10.08] \\
261 & 0.189  & [-2.469, 2.846] && 3.729 & [-1.571, 9.029] && 5.268 & [-0.564, 11.10] \\
326 & 0.387  & [-2.432, 3.205] && 4.317 & [-1.300, 9.933] && 5.900 & [-0.293, 12.09] \\
391 & 0.711  & [-2.076, 3.497] && 4.158 & [-1.468, 9.783] && 5.795 & [-0.621, 12.21] \\
456 & 0.878  & [-1.963, 3.720] && 4.341 & [-1.560, 10.24] && 5.906 & [-0.690, 12.50] \\
522 & 1.157  & [-1.810, 4.124] && 4.307 & [-1.699, 10.31] && 5.976 & [-0.731, 12.68] \\
587 & 1.370  & [-1.508, 4.247] && 4.222 & [-1.998, 10.44] && 5.804 & [-1.338, 12.95] \\
\end{tabular}
\label{tab:example2a2}
\end{table}

\section{Discussion}
In this paper, we propose a new bootstrap method for making statistical inferences on summary statistics obtained from cross-validation. We clarify the population parameter cross-validation estimates. 
More importantly, the proposed method substantially reduces the computational demands of conventional bootstrap by fitting a random effects model. Our approach complements the work of \citet{bates2021cross}, which focuses on constructing CIs for the random quantity $Err(D_n)$.

There is still a significant gap between the empirical performance of the proposed inferences in finite samples and its theoretical justification requiring large sample approximations and root $n$ regular estimates for all relevant parameters. Our simulation study shows that the distribution of the cross-validated estimate from lasso regularized regression models or random forest 
is still reasonably Gaussian, and the associated bootstrap CI performs well.  The current theoretical justification cannot cover these settings. 
Further research in this direction is warranted.

\section{Acknowledgment*}
Dr. Cai and Dr. Tian's research is supported by 5R01HL089778 from National Institutes of Health. 
Dr. Guo's research was partially supported by a grant from Research Grants Council of the Hong Kong Special Administrative Region, China (HKUST 26308323), the Seed fund of the Big Data for Bio-Intelligence Laboratory (Z0428) and the grant L0438 from the Hong Kong University of Science and Technology

%


\begin{supplement}
\stitle{Supplementary Material}
\sdescription{The Supplementary Material contains the theoretical justification, additional simulation studies and applications.}
\end{supplement}
\begin{supplement}
\stitle{Code and Data}
\sdescription{The Code and Data contains the codes and dataset connected to this paper.}
\end{supplement}


\bibliographystyle{imsart-nameyear} 
\bibliography{main}       

\begin{thebibliography}{23}

\bibitem[\protect\citeauthoryear{Austern and Zhou}{2020}]{austern2020asymptotics}
\begin{barticle}[author]
\bauthor{\bsnm{Austern},~\bfnm{Morgane}\binits{M.}} \AND \bauthor{\bsnm{Zhou},~\bfnm{Wenda}\binits{W.}}
(\byear{2020}).
\btitle{Asymptotics of cross-validation}.
\bjournal{arXiv preprint arXiv:2001.11111}.
\end{barticle}
\endbibitem

\bibitem[\protect\citeauthoryear{Bates, Hastie and Tibshirani}{2021}]{bates2021cross}
\begin{barticle}[author]
\bauthor{\bsnm{Bates},~\bfnm{Stephen}\binits{S.}}, \bauthor{\bsnm{Hastie},~\bfnm{Trevor}\binits{T.}} \AND \bauthor{\bsnm{Tibshirani},~\bfnm{Robert}\binits{R.}}
(\byear{2021}).
\btitle{Cross-validation: what does it estimate and how well does it do it?}
\bjournal{arXiv preprint arXiv:2104.00673}.
\end{barticle}
\endbibitem

\bibitem[\protect\citeauthoryear{Bayle et~al.}{2020}]{bayle2020cross}
\begin{barticle}[author]
\bauthor{\bsnm{Bayle},~\bfnm{Pierre}\binits{P.}}, \bauthor{\bsnm{Bayle},~\bfnm{Alexandre}\binits{A.}}, \bauthor{\bsnm{Janson},~\bfnm{Lucas}\binits{L.}} \AND \bauthor{\bsnm{Mackey},~\bfnm{Lester}\binits{L.}}
(\byear{2020}).
\btitle{Cross-validation confidence intervals for test error}.
\bjournal{Advances in Neural Information Processing Systems}
\bvolume{33}
\bpages{16339--16350}.
\end{barticle}
\endbibitem

\bibitem[\protect\citeauthoryear{Boardman}{1974}]{boardman1974confidence}
\begin{barticle}[author]
\bauthor{\bsnm{Boardman},~\bfnm{Thomas~J}\binits{T.~J.}}
(\byear{1974}).
\btitle{Confidence intervals for variance components--a comparative Monte Carlo study}.
\bjournal{Biometrics}
\bpages{251--262}.
\end{barticle}
\endbibitem

\bibitem[\protect\citeauthoryear{Cai et~al.}{2025}]{cai2025bootstrapping}
\begin{barticle}[author]
\bauthor{\bsnm{Cai},~\bfnm{Bryan}\binits{B.}}, \bauthor{\bsnm{Luo},~\bfnm{Yuanhui}\binits{Y.}}, \bauthor{\bsnm{Guo},~\bfnm{Xinzhou}\binits{X.}}, \bauthor{\bsnm{Pellegrini},~\bfnm{Fabio}\binits{F.}}, \bauthor{\bsnm{Pang},~\bfnm{Menglan}\binits{M.}}, \bauthor{\bparticle{de} \bsnm{Moor},~\bfnm{Carl}\binits{C.}}, \bauthor{\bsnm{Shen},~\bfnm{Changyu}\binits{C.}}, \bauthor{\bsnm{Charu},~\bfnm{Vivek}\binits{V.}} \AND \bauthor{\bsnm{Tian},~\bfnm{Lu}\binits{L.}}
(\byear{2025}).
\btitle{Supplementary Material to "Bootstrapping the Cross-Validation Estimate"}.
\bjournal{Annals of Applied Statistics}.
\end{barticle}
\endbibitem

\bibitem[\protect\citeauthoryear{Chen et~al.}{2017}]{chen2017general}
\begin{barticle}[author]
\bauthor{\bsnm{Chen},~\bfnm{Shuai}\binits{S.}}, \bauthor{\bsnm{Tian},~\bfnm{Lu}\binits{L.}}, \bauthor{\bsnm{Cai},~\bfnm{Tianxi}\binits{T.}} \AND \bauthor{\bsnm{Yu},~\bfnm{Menggang}\binits{M.}}
(\byear{2017}).
\btitle{A general statistical framework for subgroup identification and comparative treatment scoring}.
\bjournal{Biometrics}
\bvolume{73}
\bpages{1199--1209}.
\end{barticle}
\endbibitem

\bibitem[\protect\citeauthoryear{Davison and Hinkley}{1997}]{davison1997bootstrap}
\begin{bbook}[author]
\bauthor{\bsnm{Davison},~\bfnm{A.~C.}\binits{A.~C.}} \AND \bauthor{\bsnm{Hinkley},~\bfnm{D.~V.}\binits{D.~V.}}
(\byear{1997}).
\btitle{Bootstrap Methods and their Application}.
\bseries{Cambridge Series in Statistical and Probabilistic Mathematics}.
\bpublisher{Cambridge University Press}.
\end{bbook}
\endbibitem

\bibitem[\protect\citeauthoryear{Dudoit and van~der Laan}{2005}]{dudoit2005asymptotics}
\begin{barticle}[author]
\bauthor{\bsnm{Dudoit},~\bfnm{Sandrine}\binits{S.}} \AND \bauthor{\bparticle{van~der} \bsnm{Laan},~\bfnm{Mark~J}\binits{M.~J.}}
(\byear{2005}).
\btitle{Asymptotics of cross-validated risk estimation in estimator selection and performance assessment}.
\bjournal{Statistical methodology}
\bvolume{2}
\bpages{131--154}.
\end{barticle}
\endbibitem

\bibitem[\protect\citeauthoryear{Efron and Tibshirani}{1994}]{efron1994introduction}
\begin{bbook}[author]
\bauthor{\bsnm{Efron},~\bfnm{Bradley}\binits{B.}} \AND \bauthor{\bsnm{Tibshirani},~\bfnm{Robert~J}\binits{R.~J.}}
(\byear{1994}).
\btitle{An introduction to the bootstrap}.
\bpublisher{CRC press}.
\end{bbook}
\endbibitem

\bibitem[\protect\citeauthoryear{Efron and Tibshirani}{1997}]{efron1997improvements}
\begin{barticle}[author]
\bauthor{\bsnm{Efron},~\bfnm{Bradley}\binits{B.}} \AND \bauthor{\bsnm{Tibshirani},~\bfnm{Robert}\binits{R.}}
(\byear{1997}).
\btitle{Improvements on cross-validation: the 632+ bootstrap method}.
\bjournal{Journal of the American Statistical Association}
\bvolume{92}
\bpages{548--560}.
\end{barticle}
\endbibitem

\bibitem[\protect\citeauthoryear{Hemann, Bimson and Taylor}{2007}]{hemann2007framingham}
\begin{barticle}[author]
\bauthor{\bsnm{Hemann},~\bfnm{Brian~A}\binits{B.~A.}}, \bauthor{\bsnm{Bimson},~\bfnm{William~F}\binits{W.~F.}} \AND \bauthor{\bsnm{Taylor},~\bfnm{Allen~J}\binits{A.~J.}}
(\byear{2007}).
\btitle{The Framingham Risk Score: an appraisal of its benefits and limitations}.
\bjournal{American Heart Hospital Journal}
\bvolume{5}
\bpages{91--96}.
\end{barticle}
\endbibitem

\bibitem[\protect\citeauthoryear{Investigators}{2004}]{peace2004angiotensin}
\begin{barticle}[author]
\bauthor{\bsnm{Investigators},~\bfnm{PEACE~Trial}\binits{P.~T.}}
(\byear{2004}).
\btitle{Angiotensin-converting--enzyme inhibition in stable coronary artery disease}.
\bjournal{New England Journal of Medicine}
\bvolume{351}
\bpages{2058--2068}.
\end{barticle}
\endbibitem

\bibitem[\protect\citeauthoryear{Krittanawong et~al.}{2017}]{krittanawong2017artificial}
\begin{barticle}[author]
\bauthor{\bsnm{Krittanawong},~\bfnm{Chayakrit}\binits{C.}}, \bauthor{\bsnm{Zhang},~\bfnm{HongJu}\binits{H.}}, \bauthor{\bsnm{Wang},~\bfnm{Zhen}\binits{Z.}}, \bauthor{\bsnm{Aydar},~\bfnm{Mehmet}\binits{M.}} \AND \bauthor{\bsnm{Kitai},~\bfnm{Takeshi}\binits{T.}}
(\byear{2017}).
\btitle{Artificial intelligence in precision cardiovascular medicine}.
\bjournal{Journal of the American College of Cardiology}
\bvolume{69}
\bpages{2657--2664}.
\end{barticle}
\endbibitem

\bibitem[\protect\citeauthoryear{Laird and Ware}{1982}]{laird1982random}
\begin{barticle}[author]
\bauthor{\bsnm{Laird},~\bfnm{Nan~M}\binits{N.~M.}} \AND \bauthor{\bsnm{Ware},~\bfnm{James~H}\binits{J.~H.}}
(\byear{1982}).
\btitle{Random-effects models for longitudinal data}.
\bjournal{Biometrics}
\bpages{963--974}.
\end{barticle}
\endbibitem

\bibitem[\protect\citeauthoryear{LeDell, Petersen and van~der Laan}{2015}]{ledell2015computationally}
\begin{barticle}[author]
\bauthor{\bsnm{LeDell},~\bfnm{Erin}\binits{E.}}, \bauthor{\bsnm{Petersen},~\bfnm{Maya}\binits{M.}} \AND \bauthor{\bparticle{van~der} \bsnm{Laan},~\bfnm{Mark}\binits{M.}}
(\byear{2015}).
\btitle{Computationally efficient confidence intervals for cross-validated area under the ROC curve estimates}.
\bjournal{Electronic journal of statistics}
\bvolume{9}
\bpages{1583}.
\end{barticle}
\endbibitem

\bibitem[\protect\citeauthoryear{Lei}{2020}]{lei2020cross}
\begin{barticle}[author]
\bauthor{\bsnm{Lei},~\bfnm{Jing}\binits{J.}}
(\byear{2020}).
\btitle{Cross-validation with confidence}.
\bjournal{Journal of the American Statistical Association}
\bvolume{115}
\bpages{1978--1997}.
\end{barticle}
\endbibitem

\bibitem[\protect\citeauthoryear{Solomon et~al.}{2006}]{solomon2006renal}
\begin{barticle}[author]
\bauthor{\bsnm{Solomon},~\bfnm{Scott~D}\binits{S.~D.}}, \bauthor{\bsnm{Rice},~\bfnm{Madeline~M}\binits{M.~M.}}, \bauthor{\bsnm{A.~Jablonski},~\bfnm{Kathleen}\binits{K.}}, \bauthor{\bsnm{Jose},~\bfnm{Powell}\binits{P.}}, \bauthor{\bsnm{Domanski},~\bfnm{Michael}\binits{M.}}, \bauthor{\bsnm{Sabatine},~\bfnm{Marc}\binits{M.}}, \bauthor{\bsnm{Gersh},~\bfnm{Bernard~J}\binits{B.~J.}}, \bauthor{\bsnm{Rouleau},~\bfnm{Jean}\binits{J.}}, \bauthor{\bsnm{Pfeffer},~\bfnm{Marc~A}\binits{M.~A.}} \AND \bauthor{\bsnm{Braunwald},~\bfnm{Eugene}\binits{E.}}
(\byear{2006}).
\btitle{Renal function and effectiveness of angiotensin-converting enzyme inhibitor therapy in patients with chronic stable coronary disease in the Prevention of Events with ACE inhibition (PEACE) trial}.
\bjournal{Circulation}
\bvolume{114}
\bpages{26--31}.
\end{barticle}
\endbibitem

\bibitem[\protect\citeauthoryear{Sullivan, Massaro and D'Agostino~Sr}{2004}]{sullivan2004presentation}
\begin{barticle}[author]
\bauthor{\bsnm{Sullivan},~\bfnm{Lisa~M}\binits{L.~M.}}, \bauthor{\bsnm{Massaro},~\bfnm{Joseph~M}\binits{J.~M.}} \AND \bauthor{\bsnm{D'Agostino~Sr},~\bfnm{Ralph~B}\binits{R.~B.}}
(\byear{2004}).
\btitle{Presentation of multivariate data for clinical use: The Framingham Study risk score functions}.
\bjournal{Statistics in medicine}
\bvolume{23}
\bpages{1631--1660}.
\end{barticle}
\endbibitem

\bibitem[\protect\citeauthoryear{Tian et~al.}{2007}]{tian2007model}
\begin{barticle}[author]
\bauthor{\bsnm{Tian},~\bfnm{Lu}\binits{L.}}, \bauthor{\bsnm{Cai},~\bfnm{Tianxi}\binits{T.}}, \bauthor{\bsnm{Goetghebeur},~\bfnm{Els}\binits{E.}} \AND \bauthor{\bsnm{Wei},~\bfnm{LJ}\binits{L.}}
(\byear{2007}).
\btitle{Model evaluation based on the sampling distribution of estimated absolute prediction error}.
\bjournal{Biometrika}
\bvolume{94}
\bpages{297--311}.
\end{barticle}
\endbibitem

\bibitem[\protect\citeauthoryear{Tian et~al.}{2014}]{tian2014simple}
\begin{barticle}[author]
\bauthor{\bsnm{Tian},~\bfnm{Lu}\binits{L.}}, \bauthor{\bsnm{Alizadeh},~\bfnm{Ash~A}\binits{A.~A.}}, \bauthor{\bsnm{Gentles},~\bfnm{Andrew~J}\binits{A.~J.}} \AND \bauthor{\bsnm{Tibshirani},~\bfnm{Robert}\binits{R.}}
(\byear{2014}).
\btitle{A simple method for estimating interactions between a treatment and a large number of covariates}.
\bjournal{Journal of the American Statistical Association}
\bvolume{109}
\bpages{1517--1532}.
\end{barticle}
\endbibitem

\bibitem[\protect\citeauthoryear{Williams}{1962}]{williams1962confidence}
\begin{barticle}[author]
\bauthor{\bsnm{Williams},~\bfnm{JS}\binits{J.}}
(\byear{1962}).
\btitle{A confidence interval for variance components}.
\bjournal{Biometrika}
\bvolume{49}
\bpages{278--281}.
\end{barticle}
\endbibitem

\bibitem[\protect\citeauthoryear{Yadlowsky et~al.}{2021}]{yadlowsky2021estimation}
\begin{barticle}[author]
\bauthor{\bsnm{Yadlowsky},~\bfnm{Steve}\binits{S.}}, \bauthor{\bsnm{Pellegrini},~\bfnm{Fabio}\binits{F.}}, \bauthor{\bsnm{Lionetto},~\bfnm{Federica}\binits{F.}}, \bauthor{\bsnm{Braune},~\bfnm{Stefan}\binits{S.}} \AND \bauthor{\bsnm{Tian},~\bfnm{Lu}\binits{L.}}
(\byear{2021}).
\btitle{Estimation and validation of ratio-based conditional average treatment effects using observational data}.
\bjournal{Journal of the American Statistical Association}
\bvolume{116}
\bpages{335--352}.
\end{barticle}
\endbibitem

\bibitem[\protect\citeauthoryear{Yousef}{2021}]{yousef2021estimating}
\begin{barticle}[author]
\bauthor{\bsnm{Yousef},~\bfnm{Waleed~A}\binits{W.~A.}}
(\byear{2021}).
\btitle{Estimating the standard error of cross-Validation-Based estimators of classifier performance}.
\bjournal{Pattern Recognition Letters}
\bvolume{146}
\bpages{115--125}.
\end{barticle}
\endbibitem

\end{thebibliography}


\end{document}